\documentclass{aa}

\usepackage{graphicx}
\usepackage{natbib}
\usepackage{txfonts}
\usepackage{aalongtable}
\bibpunct{(}{)}{;}{a}{}{,}

\begin{document}

\title{Accurate Ritz wavelengths of parity-forbidden [\ion{Fe}{ii}], [\ion{Ti}{ii}] and [\ion{Cr}{ii}] infrared lines of astrophysical interest}

\author{M. Aldenius
\and S. Johansson}


\institute{Atomic Astrophysics, Lund Observatory, Lund University, Box 43, SE-221 00 Lund, Sweden\\
\email{maria@astro.lu.se, sveneric.johansson@astro.lu.se}}

\date{Received 16 January 2007 / Accepted 6 March 2007}

\authorrunning{Aldenius and Johansson}
\titlerunning{Ritz wavelengths of forbidden lines}


\abstract
{With new astronomical infrared spectrographs the demands of accurate atomic data in the infrared have increased. In this region there is a large amount of parity-forbidden lines, which are of importance in diagnostics of low-density astrophysical plasmas.} 
{We present improved, experimentally determined, energy levels for the lowest even LS terms of \ion{Fe}{ii}, \ion{Ti}{ii} and \ion{Cr}{ii}, along with accurate Ritz wavelengths for parity-forbidden transitions between and within these terms.} 
{Spectra of \ion{Fe}{ii}, \ion{Ti}{ii} and \ion{Cr}{ii} have been produced in a hollow cathode discharge lamp and acquired using high-resolution Fourier Transform (FT) spectrometry. The energy levels have been determined by using observed allowed ultraviolet transitions connecting the even terms with upper odd terms. Ritz wavelengths of parity-forbidden lines have then been determined.} 
{Energy levels of the four lowest \ion{Fe}{ii} terms (a$^{6}$D, a$^{4}$F, a$^{4}$D and a$^{4}$P) have been determined, resulting in 97 different parity-forbidden transitions with wavelengths between 0.74 and 87\,$\mu$m. For \ion{Ti}{ii} the energy levels of the two lowest terms (a$^{4}$F and b$^{4}$F) have been determined, resulting in 24 different parity-forbidden transitions with wavelengths between 8.9 and 130\,$\mu$m. Also for \ion{Cr}{ii} the energy levels of the two lowest terms (a$^{6}$S and a$^{6}$D) have been determined, in this case resulting in 12 different parity-forbidden transitions with wavelengths between 0.80 and 140\,$\mu$m.} 
{} 


\maketitle



\section{Introduction}\label{Introduction_sect}

The interaction between laboratory and stellar spectroscopy has a long history, and still the availability of experimental atomic data is a necessity in analyses of stellar spectra. The development in astronomical instrumentation has also increased the demands on the accuracy of the data in the optical (e.g. the VLT/UVES spectrograph) and ultraviolet (e.g. the HST/STIS instrument) wavelength regions. The next step will be to improve the atomic database in the near-infrared region to match the needs defined by high-resolution spectra recorded with the newly installed CRIRES spectrograph at VLT and the space-borne SPITZER observatory.

The situation in nebular spectroscopy is quite different from the status of stellar spectroscopy. Nebular spectra show emission lines, and they are associated with various excitation mechanisms, general as well as selective. Very special features in nebular spectra are the "forbidden lines", i.e. transitions between metastable (long-lived) energy levels, which are only observable in spectra of very low-density plasmas. The forbidden lines are important for diagnostics of various parts of planetary nebulae \citep[e.g.][]{Smith05} and nebular clouds around extended objects, such as active galactic nuclei, \citep[see e.g.][]{Kaufman06,Meijerink07}. The diagnostics are often performed using standard pairs of forbidden lines, and the information embedded in numerous other forbidden lines is seldom exploited. This is partly due to the lack of reliable atomic data for forbidden lines.

Forbidden lines are not observed in light sources used for atomic spectroscopy, simply because collisions deexcite the metastable levels due to the high electron density. The wavelengths can be calculated from the energy level values, but it is not until recently lifetime measurements of metastable levels in complex spectra have been performed  using storage rings \citep{Hartman05}. Astrophysical spectra have been used for converting the lifetimes into Einstein $A$-coefficients. Great efforts were early made by Garstang \citep[e.g][]{Garstang62} to calculate the $A$-coefficients, and they have later been followed up by work by Nussbaumer \citep{Nussbaumer70,Nussbaumer88}, Quinet \citep{Quinet96,Quinet97} and others. The theoretical data are certainly of high quality but they are not accompanied with any estimate of the uncertainty. However, accurate wavelengths can only be derived from experimental data.

In the present paper we present new and accurate wavelengths of forbidden lines in the infrared region for three spectra, \ion{Fe}{ii}, \ion{Ti}{ii} and \ion{Cr}{ii}, of high astrophysical relevance. Many of these lines have already been identified in astrophysical spectra, see e.g. \citet{Temim06,Riffel06,Hartman04}. Other second spectra of the iron group elements produce few forbidden lines in the infrared either because of the atomic structure or the low abundance. The energy level values of the lowest terms in \ion{Fe}{ii}, \ion{Ti}{ii} and \ion{Cr}{ii} have been improved by using observed UV-visible transitions measured with high-resolution Fourier Transform spectrometry. Accurate Ritz wavelengths have been determined for parity-forbidden transitions between and within these terms.

\section{Experimental method}\label{Method_sect}

\begin{table*}
 \caption{Lamp conditions, spectrometer resolution and correction factors for the spectral acquisitions}
 \label{exp_table}
 \centering
 \begin{tabular}{ccccccrcc}
 \hline\hline
 \\[-3mm]
 Spectrum & Wavenumber & Cathode & Inserted & Carrier gas & Pressure & Current & Resolution & Calibration factor\\
 number & range (cm$^{-1}$) & material & metals & {} & (torr) & (mA) & (cm$^{-1}$) & $k_{\mathrm{eff}}$\\
 \hline
 \\[-3mm]
 I    & 22200--44200 & Ti & & Ar & 0.6 & 500 & 0.05 & $(1.43\pm 0.02)\cdot10^{-6}$\\ 
 II   & 19800--39400 & Fe & & Ar & 0.8 & 700 & 0.05 & $(-2.34\pm 0.02)\cdot10^{-6}$\\ 
 III  & 26900--53700 & Fe & & Ar & 1.0 & 700 & 0.04 & $(-2.54\pm 0.03)\cdot10^{-6}$\\ 
 IV   & 21700--43400 & Fe & & Ne & 1.2 & 1200 & 0.06 & $(-2.90\pm 0.03)\cdot10^{-6}$\\ 
 V    & 21700--43400 & Fe & & Ne & 1.0 & 900  & 0.06 & $(-2.90\pm 0.03)\cdot10^{-6}$\\ 
 VI   & 20600--41000 & Fe & Mg, Ti, Cr, Mn, Zn & Ar \& Ne & 0.9 & 700 & 0.05 & $(-2.12\pm 0.02)\cdot10^{-6}$\\ 
 VII  & 25300--50500 & Fe & Mg, Ti, Cr, Mn, Zn & Ar \& Ne & 1.0 & 600 & 0.06 & $(-2.23\pm 0.03)\cdot10^{-6}$\\ 
 VIII & 26900--53700 & Fe & Mg, Cr, Zn & Ne & 0.8 & 400 & 0.06 & $(-3.19\pm 0.04)\cdot10^{-6}$\\ 
 IX   & 26900--53700 & Fe & Mg, Cr, Zn & Ne & 0.9 & 500 & 0.06 & $(-3.13\pm 0.04)\cdot10^{-6}$\\ 
 \hline
 \end{tabular}
\end{table*}

Spectra of \ion{Fe}{ii}, \ion{Ti}{ii} and \ion{Cr}{ii} were produced in a water cooled hollow cathode discharge lamp (HCL). Four different cathode compositions were used, see Table~\ref{exp_table}. In one spectral acquisition the cathode consisted of pure titanium, spectrum I, and for the others a cathode of pure iron was used. The cathodes had a cylindrical bore with an inner diameter of 6-7\,mm and a length of about 50\,mm. For four of the spectral acquisitions pieces of complementary metals were inserted in the cathode bore to produce spectra from several different ions simultaneously. Spectra II-IX were previously acquired for measurements of accurate relative wavelengths of different ions, see \citet{Aldenius06}. These spectra were complemented with spectrum I to improve the signal-to-noise ratios (S/N) and accuracy of the \ion{Ti}{ii} lines in this investigation. For the spectral acquisition with pure Ti or Fe cathodes argon or neon was used as a carrier gas, creating the initial plasma in the cathode. This provided Ar lines for wavelength calibration and Fe lines which were used as secondary calibration lines, see Sect.~\ref{Calib_sect}. For the spectral acquisitions with composite cathodes either a mixture of argon and neon or pure neon was chosen as the carrier gas. This was done because Ne produced higher S/N for lines that were too weak in pure Ar recordings. A continuous HCL mainly produces spectra from neutral and singly ionized species. To increase the amount of singly ionized species, and the S/N for the lines of interest, a large amount of spectra were recorded to achieve the best light source parameters. Finally, the light source was run with a current between 0.4 and 1.2\,A and a pressure between 0.6 and 1.2\,torr.

The spectra were acquired with the Chelsea Instruments FT500 UV Fourier Transform (FT) spectrometer \citep{Thorne87} in Lund. This is optimized to measure high-resolution spectra in the region 2000--7000\,\AA . Spectra for essentially two overlapping wavenumber regions, 20000--40000\,cm$^{-1}$ (5000--2500\,\AA ), region A, and 26000--50000\,cm$^{-1}$ (3800--2000\,\AA ), region B, were acquired using three different Hamamatsu photomultiplier tubes, R928 or 1P28 for region A and R166 for region B. Photon noise in the interferogram is transferred into the spectral region as white noise. All spectral lines seen by the detector contribute to this noise level. It is therefore an advantage to limit the light seen by the detector to the wavelength region of interest. The R166 detector used in region B is so-called solar-blind and has a long-wavelength cut-off at about 3000\,\AA. The short-wavelength cut-off for this region is at about 2000\,\AA , where air absorbs most light. For region A the R928 detector has a wide sensitivity range, between 2000 and 8000\,\AA , and an external 2\,mm UG-5 optical glass filter was used to cut out light above 4000\,\AA , significantly reducing the noise. The two regions, including detectors and filter, overlap, giving a small region for transfer of calibration between them (see Sect.~\ref{Calib_sect}).

The light source was carefully aligned with a focusing lens at two different distances to the spectrometer. For spectra VII--IX (in region B) the light source was placed close to the aperture of the spectrometer, in order to decrease the light path length in air and thereby increase the S/N for the lines close to 2000\,\AA , where air absorbs most light. For the region at longer wavelengths (A) the light source was placed at a larger distance. In this wavelength region the absorption from air is less prominent. The resolution of the spectra was between 0.04 and 0.06\,cm$^{-1}$, which was sufficient to completely resolve the lines. For each spectrum about 20 scans were co-added to achieve higher S/N for the lines. To further improve the accuracy the allowed E1 transitions were measured in several spectral recordings with different light-source parameters.

\section{Analysis}\label{Analysis_sect}

For each of the three species the lowest LS terms of the low even configurations, including the ground term, were chosen to be included in the investigation. The upper levels of the parity-forbidden transitions should be meta-stable, having no level of opposite parity located at lower energy. It was also essential that the terms could be easily connected with UV-visible transitions through higher lying levels of opposite parity. To improve the accuracy of the lower levels, as many transitions to upper levels of odd parity as possible were chosen to be included in the measurements. Since the transition rules for allowed E1 transitions are $\Delta J=0,\pm 1$, see Table~\ref{Rules_table}, there was a maximum of three transitions to each upper level within one multiplet. To further improve the accuracy, lines were measured in at least three different spectral acquisitions.

\begin{table}
\begin{minipage}[t]{\columnwidth}
\caption{Transition rules for allowed E1 and forbidden M1 and E2 transitions}
\label{Rules_table}
\centering
\renewcommand{\footnoterule}{}  
\begin{tabular}{ccc}
\hline\hline
\\[-3mm]
E1 & M1 & E2 \\
\hline
\\[-3mm]
\multicolumn{3}{c}{Rules independent of coupling scheme} \\
\hline
parity change & no parity change & no parity change \\
$\Delta J=0,\pm 1$ & $\Delta J=0,\pm 1$ & $\Delta J=0,\pm 1,\pm 2$ \\
$(0 \nleftrightarrow 0)$ & $(0 \nleftrightarrow 0)$ & $(0 \nleftrightarrow 0, \frac{1}{2} \nleftrightarrow \frac{1}{2}, 0 \nleftrightarrow 1)$ \\
\hline
\end{tabular}
\end{minipage}
\end{table}

\begin{figure}
 \includegraphics[width=84mm]{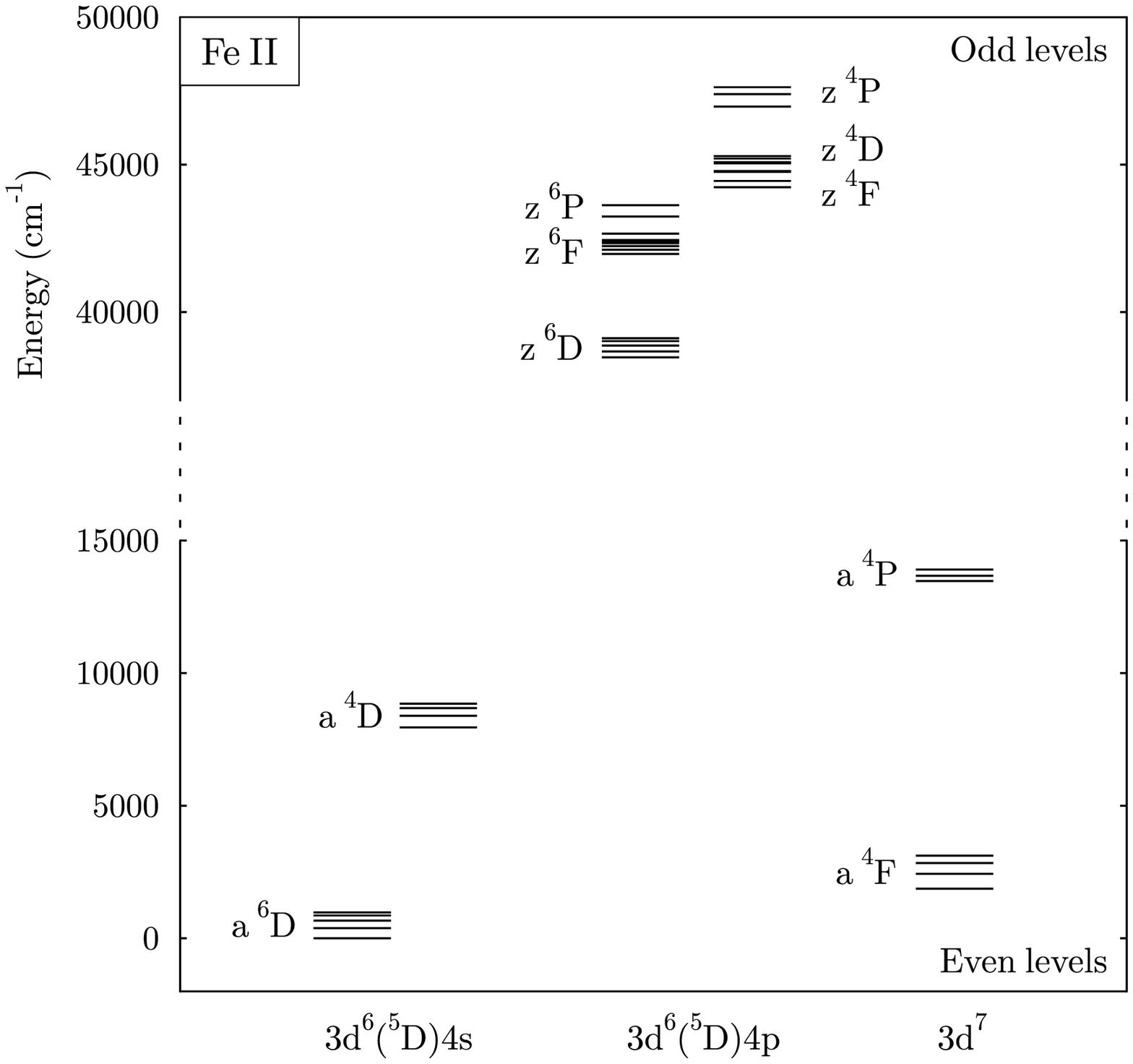}
 \caption{Partial energy level diagram of \ion{Fe}{ii} displaying the low even levels and higher odd levels included in this investigation. Parity-forbidden M1 and E2 transitions occur between, and within, the even terms, while allowed E1 transitions between the odd and even terms are used for determining accurate energy level values.}
 \label{FeIIlev_fig}
\end{figure}

\begin{figure}
 \includegraphics[width=84mm]{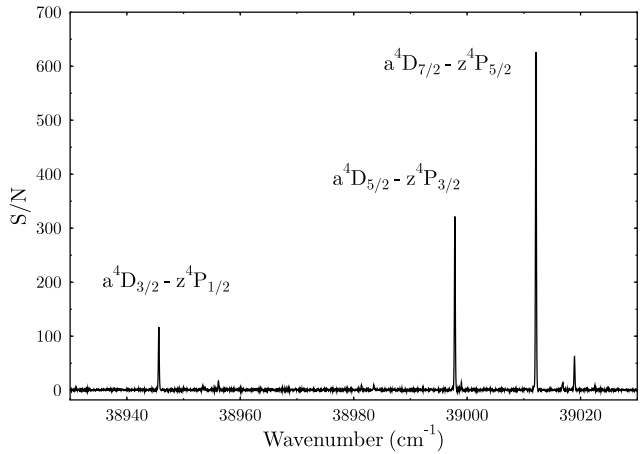}
 \caption{Observed \ion{Fe}{ii} transitions, belonging to the a$^{4}$D--z$^{4}$P multiplet, from spectrum III. The wavelength span in the figure is about 6.6\,\AA\ and the weaker line at 39019\,cm$^{-1}$ is an \ion{Ar}{ii} line from the carrier gas in the HCL.}
 \label{FeIIspec_fig}
\end{figure}

\ion{Fe}{ii} has a complex energy level structure, with many low terms within a small energy interval. The ground state is 3d$^{6}$($^{5}$D)4s\,a$^{6}$D$_{9/2}$ and the lowest terms belong to the even 3d$^{6}$($^{5}$D)4s and 3d$^{7}$ configurations. The lower, even parity, energy levels of four different terms (a$^{6}$D, a$^{4}$F, a$^{4}$D and a$^{4}$P), with in total 16 energy levels between 0 and 13900\,cm$^{-1}$, were included in the investigation. Allowed UV-visible transitions to six terms of odd parity (z$^{6}$D, z$^{6}$F, z$^{6}$P, z$^{4}$F, z$^{4}$D and z$^{4}$P) were measured. These terms all belong to the 3d$^{6}$($^{5}$D)4p configuration and are the lowest odd parity terms in \ion{Fe}{ii} with energies between 38000 and 48000\,cm$^{-1}$. A partial energy level diagram of \ion{Fe}{ii} is displayed in Fig.~\ref{FeIIlev_fig}, including the even and odd terms considered in this investigation. Because of rather poor LS-coupling in \ion{Fe}{ii}, transitions between quartets and sextets have high probabilities and such intercombination lines were in general clearly visible in the laboratory spectra. This made it possible to connect the two spin systems. Fig.~\ref{FeIIspec_fig} shows a small part of an observed spectrum (III) including three of the \ion{Fe}{ii} lines used in this investigation. The spectrum is from a pure iron cathode and contains iron and argon lines.

\begin{figure}
 \includegraphics[width=84mm]{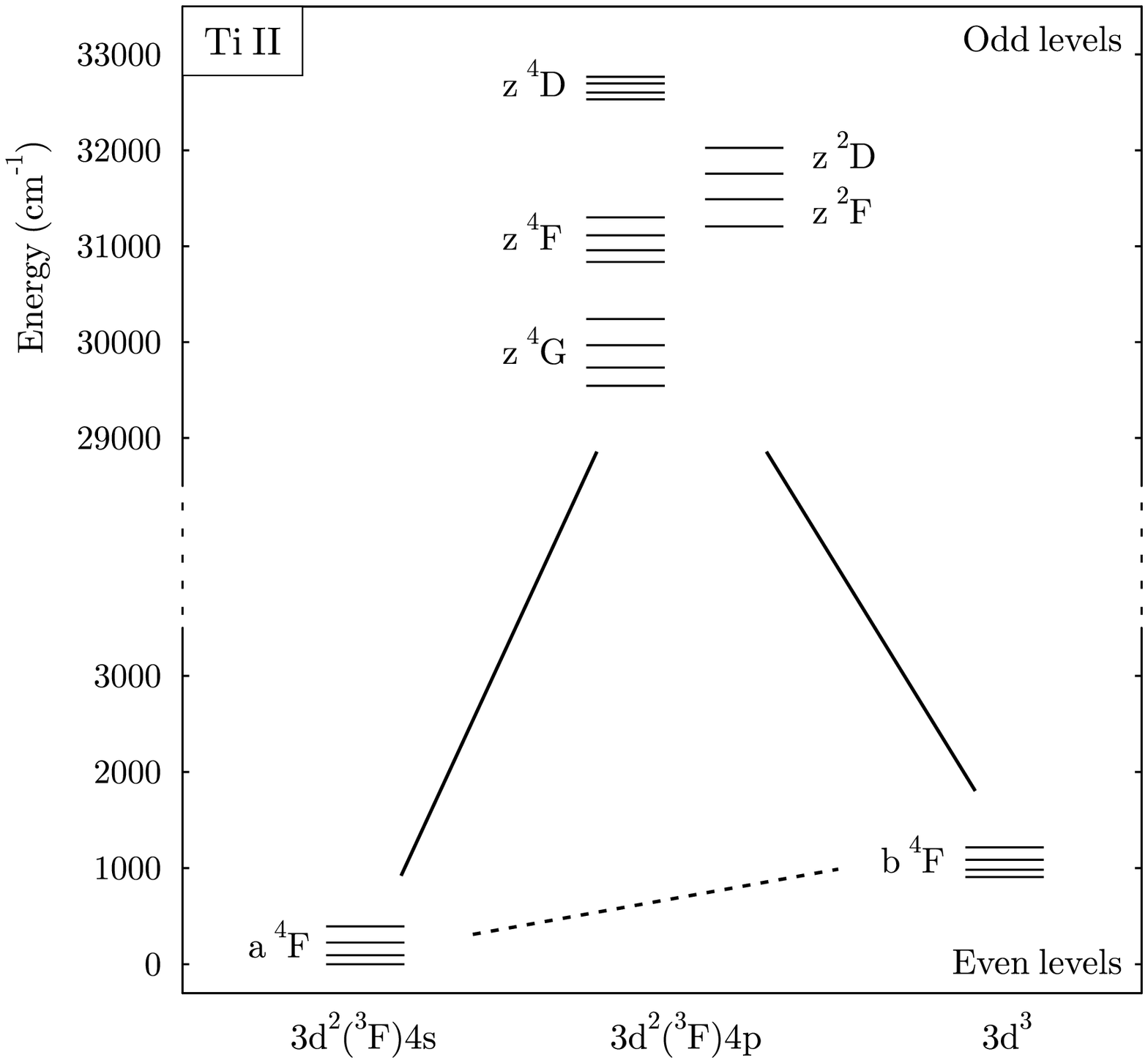}
 \caption{Partial energy level diagram of \ion{Ti}{ii} displaying the low even levels and higher odd levels included in this investigation. Parity-forbidden M1 and E2 transitions (dashed line) occur between, and within, the even terms, while allowed E1 transitions (solid lines) between the odd and even terms are used for determining accurate energy level values.}
 \label{TiIIlev_fig}
\end{figure}

Also \ion{Ti}{ii} has a complex energy level structure with many low terms of even parity. The ground state is 3d$^{2}$($^{3}$F)4s\,a$^{4}$F$_{3/2}$ and the lowest terms belong to the 3d$^{2}$($^{3}$F)4s and 3d$^{3}$ configurations. The lower, even parity, energy levels of two different terms (a$^{4}$F and b$^{4}$F), with in total 8 energy levels between 0 and 1200\,cm$^{-1}$, were included in the investigation. Allowed UV-visible transitions to five terms of odd parity (z$^{4}$G, z$^{4}$F, z$^{2}$F, z$^{2}$D and z$^{4}$D) were measured. These terms all belong to the 3d$^{2}$($^{3}$F)4p configuration and are the lowest odd parity terms in \ion{Ti}{ii} with energies between 29000 and 33000\,cm$^{-1}$. A partial energy level diagram of \ion{Ti}{ii} including the even and odd terms in this investigation is shown in Fig.~\ref{TiIIlev_fig}.

For \ion{Cr}{ii} the ground state is 3d$^{5}$\,a$^{6}$S$_{5/2}$ and the lowest terms belong to the 3d$^{5}$ and 3d$^{4}$($^{5}$D)4s configurations. The lower, even parity, energy levels of two different terms (a$^{6}$S and a$^{6}$D), with in total 6 energy levels between 0 and 12500\,cm$^{-1}$, were included in the investigation. The ground term has zero orbital angular momentum, $L=0$, and therefore only one energy level ($J=\frac{5}{2}$). This means that the total number of possible transitions within, and between, these two lower terms are less than for e.g. \ion{Ti}{ii}. The lowest odd parity term in \ion{Cr}{ii} is z$^{6}$F at 47000\,cm$^{-1}$, but transitions to the ground state from this term have very low transition probabilities \citep{Nilsson06}. The second lowest odd term, z$^{6}$P, is at 48000\,cm$^{-1}$, which results in transitions close to the 2000\,\AA\ limit were air absorbs most light, making the observed transitions with low probabilities too weak. This made it possible to measure transitions only to (z$^{6}$P), belonging to the 3d$^{4}$($^{5}$D)4p configuration. To try and compensate for this lack off useable transitions and improve the accuracies, the \ion{Cr}{ii} lines were measured in four different spectra.

\subsection{Wavenumber calibration}\label{Calib_sect}

FT spectrometry generates spectra with a linear wavenumber scale, whose accuracy derives from the control of the sampling of the interferogram by a single-mode helium-neon laser. The frequency of this control laser is stabilized to five parts in $10^{9}$. The accuracy is, however, limited by the effects of using a finite-size aperture and by imperfect alignment of the light from the light source and the control laser \citep{Learner88}. To obtain a wavenumber scale, which is accurate to better than one part in $10^{7}$, a multiplicative correction is applied, such that
\begin{equation}
\sigma_{\mathrm{corr}}=\left( 1+k_{\mathrm{eff}}\right) \sigma_{\mathrm{obs}}
\end{equation}
where $\sigma_{\mathrm{corr}}$ is the corrected wavenumber and $\sigma_{\mathrm{obs}}$ is the observed, uncorrected wavenumber. The correction factor, $k_{\mathrm{eff}}$, is accurately determined by measuring the positions of well-known internal wavenumber standard lines. With FT spectra it is not necessary that the reference lines are evenly distributed throughout the spectrum. In principle, it is possible to use only one calibration line \citep{Salit04}, but to reduce the uncertainty of the calibration, several calibration lines have been used for each spectrum. The use of internal calibration lines, in this case \ion{Ar}{ii} lines from the carrier gas in the HCL, helps to ensure, to as high degree as possible, that the light from the species being used for calibration illuminates the entrance aperture of the FT spectrometer in the same way as that of the species being investigated. \ion{Ar}{ii} lines are commonly used for wavelength calibration and there are in principle two sets of standard lines available. In this investigation wavenumbers from \citet{Whaling95} have been used for calibration, rather than the work by \citet{Norlen73}. For discussions on the difference between these two sets of lines see e.g.~\citet{Whaling02,Nave04,Aldenius06}. The wavenumbers of Whaling were measured using FT spectrometry with molecular CO lines as wavenumber standards.

For spectral region A \ion{Ar}{ii} 4s-4p calibration lines with S/N$>$100 were selected for the spectra with Ar as the carrier gas in the HCL (spectrum I, II and VI). Using only 4s-4p transitions decreases the risk of the lines being pressure shifted, since the pressure shift generally increases with excitation. 14 \ion{Ar}{ii} lines were used for each spectrum and the corresponding correction factor was derived by calculating the weighted mean of the 14 individual correction factors, where the weights were scaled with the inverse variance of each correction factor.

In spectral region B (spectrum III, VII, VIII and IX), covering higher wavenumbers, there are an insufficient number of reliable \ion{Ar}{ii} 4s-4p calibration lines available in our spectra. A transfer of calibration between the two spectral regions was therefore used. In spectrum VI, which includes the 14 \ion{Ar}{ii} calibration lines, a number of \ion{Fe}{i} and \ion{Fe}{ii} lines were measured and calibrated. These lines are located in the overlap of the two spectral regions and are observed in all spectra containing iron. They could therefore be used, together with three \ion{Ar}{ii} lines when present, for calibration in region B. These Fe lines were also used for calibration in the spectra in region A with pure Ne as the carrier gas (spectrum IV and V). The uncertainty of the calibration in these spectra is slightly larger due to this two-step calibration, see Table~\ref{exp_table}.

\subsection{Line profiles and blends}\label{Fit_sect}

The observed lines were fitted with Voigt profiles using a least-square procedure included in the FT spectrometry analysis program {\sc Xgremlin} \citep{Nave97b}, which is based on the {\sc Gremlin} code \citep{Brault89}. This program was also used for transforming and phase correcting the measured interferograms. The fitted Voigt profiles showed low damping constants, implying that the observed lines were close to pure Gaussians. This was expected, since Doppler broadening is the dominant broadening effect in the HCL.

All lines, including calibration lines, had an apparently symmetric peak profile and showed no form of isotope or hyperfine structure. The metals all consisted of terrestrial relative abundances of the different isotopes. For Cr and Fe any isotope or hyperfine structure should be negligible compared to the Doppler width ($\sim$0.1--0.2\,cm$^{-1}$) and they both have one dominating even isotope ($^{52}\mathrm{Cr}$:\,84\,\% and $^{56}\mathrm{Fe}$:\,92\,\%). Ti consists of one even isotope with large relative abundance ($^{48}\mathrm{Ti}$:\,74\,\%) and four other isotopes with relative abundances between 5 and 8\,\%. Unresolved isotope structure itself should not pose a problem in wavenumber measurements, but if the observed lines are self-absorbed the different isotopes will be affected by varying amount of self-absorption and a small shift in the observed wavenumber could be introduced. If self-absorption was detected or suspected in a line this was investigated more thoroughly. By changing the discharge current in the HCL, and thereby producing different plasma densities, spectra were acquired with different amount of self-absorption. No tendency of wavenumber shift or asymmetric profiles were detected. All lines in the investigation were also measured in at least three different spectra with different plasma densities, and the possible effects should be negligible compared to the uncertainties from calibration and line fitting, see Sect.~\ref{Uncert_sect}. Furthermore, all energy levels were derived by using several different transitions and observed lines which clearly gave different energy levels were removed from the fitting of energy levels, see Sect.~\ref{Efit_sect}.

All lines used in this work were carefully checked for possible blends from both the same species and other species present in the HCL. By using different HCL conditions with different carrier gas, cathode material and inserted metals the risk of not detecting unresolved blends was minimized.

Pressure shifts affect higher levels the most. In this investigation the highest levels included is at 48000\,cm$^{-1}$, which is relatively low. \citet{Nave97a} estimated a shift of less than 0.005\,cm$^{-1}$ for levels at 100000\,cm$^{-1}$ in \ion{Fe}{ii} when using a carrier gas pressure of at least 4 times higher than in this investigation. The spectra in this work were also tested for pressure shifts, see~\citet{Aldenius06}, but no shifts were noted within 0.8 to 1.2\,torr. In this investigation the difference in wavenumber between allowed transitions is used to determine the energy of the metastable levels and thus possible effects of pressure shifts on the higher odd levels become negligible.

\subsection{Fitting of energy levels}\label{Efit_sect}

\begin{figure}
 \includegraphics[width=84mm]{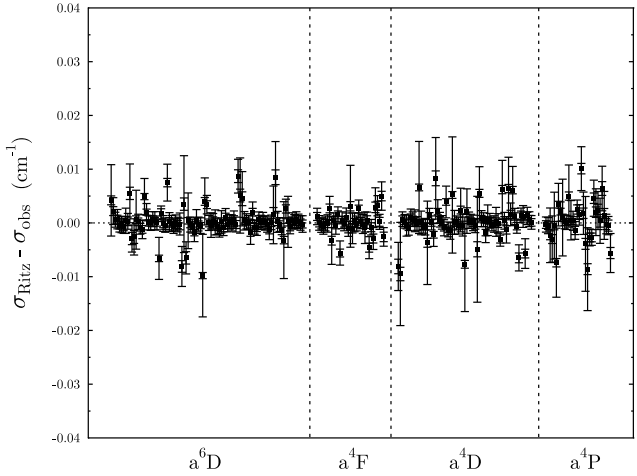}
 \caption{The difference between calculated Ritz wavenumbers and observed wavenumbers for the observed UV \ion{Fe}{ii} lines. The x-axis is ordered in the four different lower terms and increases with the energy of the lower level and the observed wavenumber of the transition. The larger and smaller error bars represent the uncertainties in the observed wavenumbers and Ritz wavenumbers, respectively.}
 \label{FeIIlevfit_fig}
\end{figure}

For each species an energy level system was simultaneously fitted to all observed transitions, using a weighted least-square method. The weight of the wavenumber of an observed transition was based on the inverse variance and the ground state was considered fixed at zero energy with zero uncertainty. The procedure is described in detail in \citet{Oberg07} and is based on the method proposed by \citet{Radziemski72}. Ritz wavenumbers could then be derived from the fitted energy levels and the difference between these and the observed wavenumbers could be studied, see Fig.~\ref{FeIIlevfit_fig} for \ion{Fe}{ii}. Observed transitions which clearly deviated from the Ritz wavenumber ($|\sigma_{\mathrm{Ritz}}-\sigma_{\mathrm{obs}}|>1.5s_{\mathrm{obs}}$) where investigated more thoroughly. If they showed any asymmetry in the observed profile or had a signal-to-noise ratio below 10 they were removed from the fitting procedure and the energy system was re-analyzed. This was done to avoid the influence from possible unknown unresolved blends, which could shift the wavenumber of the observed transition. Since all transitions were observed in several spectra, with different light source parameters, there were enough transitions to accurately determine the energy levels. For the cases of a transition being measured in several spectral acquisitions all of the observed wavenumbers were included in the fit of energy levels instead of first calculating a weighted mean.

For \ion{Fe}{ii} 96 different allowed transitions were used for determining the energy levels. The majority of the transitions were observed in at least two spectral recordings, and therefore in total 223 observed spectral lines, ranging from 31000 to 44400\,cm$^{-1}$, were used in the analysis of \ion{Fe}{ii}. The transitions to the a$^{4}$P term have comparably low transition probabilities (see e.g. \citealt{Kurucz95}) and therefore appear as weak (S/N$\approx$5--50), or not detectable, lines in the spectra acquired with Ar as the carrier gas in the HCL. To compensate for this recordings with pure Ne as the carrier gas were used, to increase the S/N of the observed lines. The values of the a$^{4}$P levels are therefore only slightly less accurate than the energies belonging to the lowest terms (a$^{6}$D, a$^{4}$F and a$^{4}$D). In Fig.~\ref{FeIIlevfit_fig} the difference between the calculated Ritz wavenumbers and the measured wavenumbers for the 223 observed lines is shown. The figure is divided into four parts showing transitions to the four different lower terms. The transitions to the three lowest terms show mostly small error bars and are closely placed around zero difference, while the transitions to a$^{4}$P show a larger variance and a higher relative number of transitions with larger error bars. This is due to the weaker observed transitions to this term. The final values of the \ion{Fe}{ii} energy levels are presented in  Table~\ref{Fe_levels_table}, together with the uncertainties and the number of observed spectral features used for determining each energy level.

\begin{figure}
 \includegraphics[width=84mm]{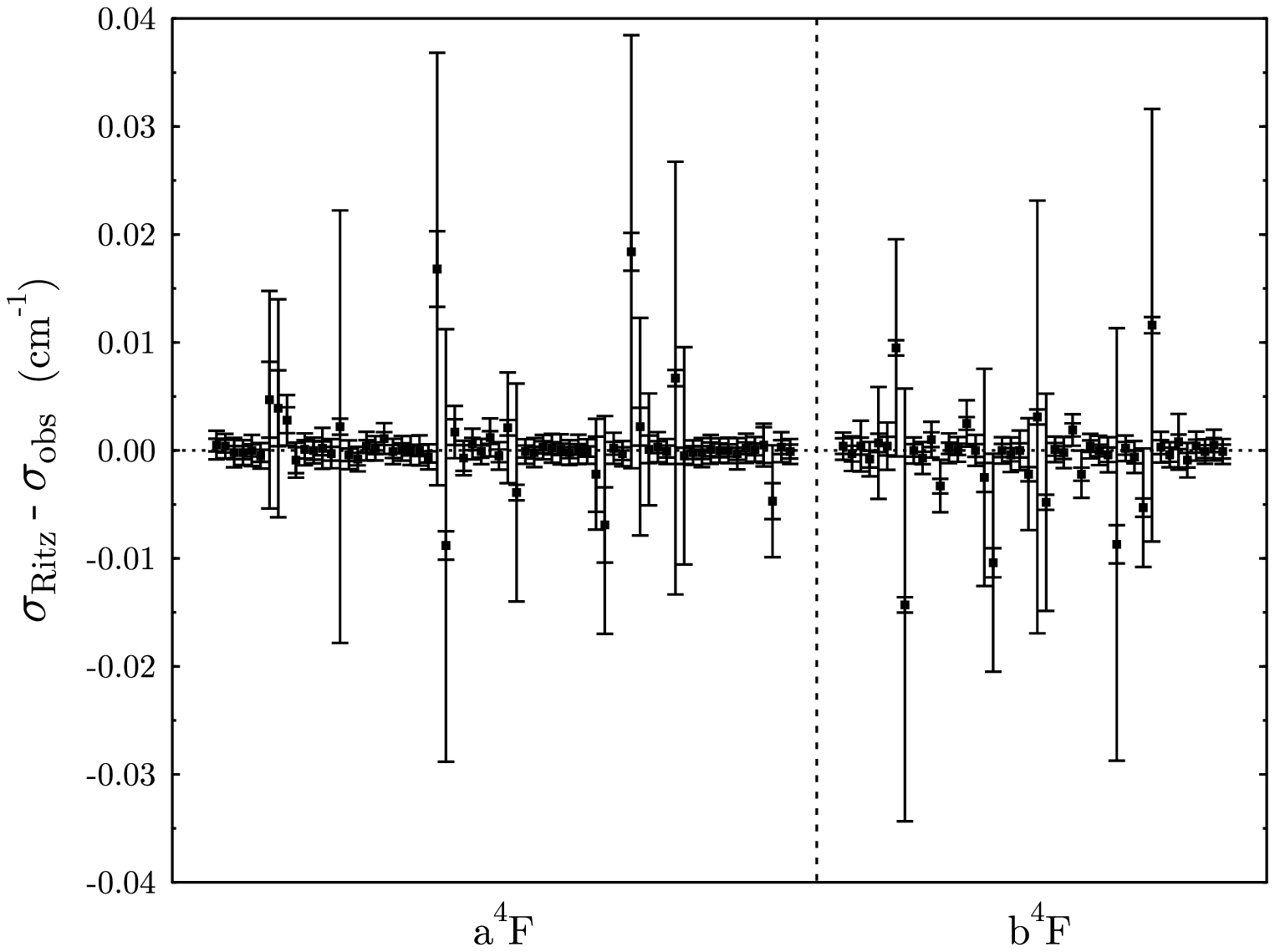}
 \caption{The difference between calculated Ritz wavenumbers and observed wavenumbers for the observed UV-visible \ion{Ti}{ii} lines. See Fig.~\ref{FeIIlevfit_fig} for description of x-axis and error bars.}
 \label{TiIIlevfit_fig}
\end{figure}

\begin{figure}
 \includegraphics[width=84mm]{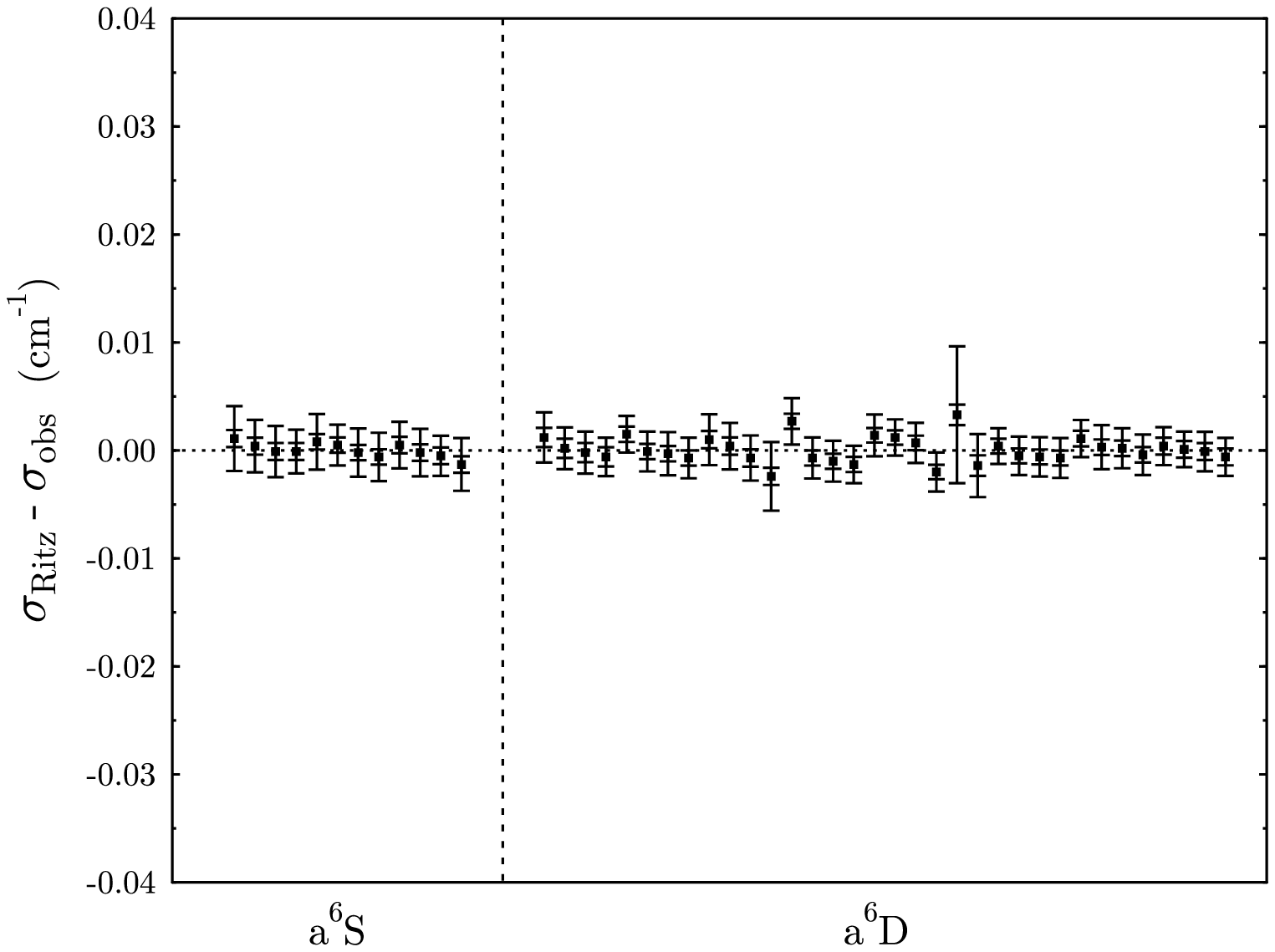}
 \caption{The difference between calculated Ritz wavenumbers and observed wavenumbers for the observed UV-visible \ion{Cr}{ii} lines. See Fig.~\ref{FeIIlevfit_fig} for description of x-axis and error bars.}
 \label{CrIIlevfit_fig}
\end{figure}

\begin{table}
\begin{minipage}[t]{\columnwidth}
\caption{Low lying \ion{Fe}{ii} energy levels of even parity}
\label{Fe_levels_table}
\centering
\renewcommand{\footnoterule}{}  
\begin{tabular}{l c c r r r}
\hline\hline
\\[-3mm]
Config. & Term & $J$ & Energy & Unc. & Lines\footnote{Number of observed spectral features used in the determination of the energy}\\
 & & & (cm$^{-1}$) & (cm$^{-1}$) & (\#) \\
\hline
\\[-3mm]
3d$^{6}$($^{5}$D)4s & a$^{6}$D & $9/2$ & 0 & 0 & 12\footnote{Energy set to zero by definition} \\
 & & $7/2$ & 384.7868 & 0.0007 & 25 \\
 & & $5/2$ & 667.6830 & 0.0008 & 27 \\
 & & $3/2$ & 862.6108 & 0.0009 & 16 \\
 & & $1/2$ & 977.0489 & 0.0010 & 13 \\
3d$^{7}$ & a$^{4}$F & $9/2$ & 1872.6005 & 0.0013 & 6 \\
 & & $7/2$ & 2430.1369 & 0.0013 & 10 \\
 & & $5/2$ & 2837.9803 & 0.0014 & 10 \\
 & & $3/2$ & 3117.4889 & 0.0015 & 7 \\
3d$^{6}$($^{5}$D)4s & a$^{4}$D & $7/2$ & 7955.3189 & 0.0012 & 18 \\
 & & $5/2$ & 8391.9562 & 0.0012 & 20 \\
 & & $3/2$ & 8680.4708 & 0.0013 & 20 \\
 & & $1/2$ & 8846.7822 & 0.0015 & 7 \\
3d$^{7}$ & a$^{4}$P & $5/2$ & 13474.4487 & 0.0013 & 15 \\
 & & $3/2$ & 13673.2019 & 0.0014 & 10 \\
 & & $1/2$ & 13904.8598 & 0.0016 & 7 \\
\hline
\end{tabular}
\end{minipage}
\end{table}

For \ion{Ti}{ii} 65 different allowed transitions were used for determining the energy levels. The majority of these transitions were observed in at least two spectral recordings, and therefore in total 110 observed spectral lines, ranging from 28500 to 39600\,cm$^{-1}$, were used in the analysis of \ion{Ti}{ii}. The two lowest terms in \ion{Ti}{ii} (a$^{4}$F and b$^{4}$F) are close in energy, within 1200\,cm$^{-1}$, but they belong to different configurations (3d$^{2}$($^{3}$F)4s and 3d$^{3}$ respectively). This means that the corresponding transitions to the two terms are close in wavenumbers and the intensity response in the spectra should be similar. Thus, the difference in observed line strength depend mainly on the difference in transition probabilities to the different terms. The transitions (within the stated wavenumber interval) to the ground term have in general higher branching fractions (and thus higher transition probabilities) \citep{Pickering01}, but in most cases both kinds of transitions were clearly visible with high S/N ($>$100). This resulted in energy levels with similar uncertainties for both lower terms. In Fig.~\ref{TiIIlevfit_fig} the differences between the calculated Ritz wavenumbers and the observed wavenumbers for the 110 measured lines are shown. The figure is divided into two parts showing transitions to the two different lower terms. The transitions to both lower terms show mostly small error bars and are closely placed around zero difference. For the lowest term (a$^{4}$F) there are slightly more transitions and the relative number of observed wavenumbers with low uncertainty is also slightly higher. This difference has, however, a small impact on the uncertainties of the energy levels. The final values of the \ion{Ti}{ii} energy levels are presented in  Table~\ref{Ti_levels_table}, together with the uncertainties and the number of observed spectral features used for determining each energy level.

\begin{table}
\begin{minipage}[t]{\columnwidth}
\caption{Low lying \ion{Ti}{ii} energy levels of even parity}
\label{Ti_levels_table}
\centering
\renewcommand{\footnoterule}{}  
\begin{tabular}{l c c r r r}
\hline\hline
\\[-3mm]
Config. & Term & $J$ & Energy & Unc. & Lines\footnote{Number of observed spectral features used in the determination of the energy}\\
 & & & (cm$^{-1}$) & (cm$^{-1}$) & (\#) \\
\hline
\\[-3mm]
3d$^{2}$($^{3}$F)4s & a$^{4}$F & $3/2$ & 0 & 0 & 15\footnote{Energy set to zero by definition} \\
 & & $5/2$ & 94.1105 & 0.0006 & 18 \\
 & & $7/2$ & 225.7014 & 0.0007 & 19 \\
 & & $9/2$ & 393.4442 & 0.0008 & 14 \\
3d$^{3}$ & b$^{4}$F & $3/2$ & 907.9646 & 0.0008 & 6 \\
 & & $5/2$ & 983.9127 & 0.0006 & 16 \\
 & & $7/2$ & 1087.3530 & 0.0007 & 13 \\
 & & $9/2$ & 1215.8304 & 0.0009 & 9 \\
\hline
\end{tabular}
\end{minipage}
\end{table}

For \ion{Cr}{ii} 12 different allowed transitions were used for determining the energy levels. The majority of these transitions were observed in at least two spectral recordings, and therefore in total 46 observed spectral lines, ranging from 36100 to 48600\,cm$^{-1}$, were used in the analysis of \ion{Cr}{ii}. As discussed in the introduction to Sect.~\ref{Analysis_sect} only transitions from one upper term was used for the determination of the energy levels of the two lowest terms in \ion{Cr}{ii}, but since four different acquired spectra were used the minimum number of observed lines used to determine an energy level was four. However, the uncertainties for the \ion{Cr}{ii} energy levels are somewhat higher than for most lower levels in \ion{Fe}{ii} and \ion{Ti}{ii}, due to the lack of transitions. In Fig.~\ref{CrIIlevfit_fig} the difference between the calculated Ritz wavenumbers and the observed wavenumbers for the 46 measured lines is shown. The figure is divided into two parts showing transitions to the two different lower terms. As the number of different transitions is much lower in this figure than in Fig.~\ref{FeIIlevfit_fig} and \ref{TiIIlevfit_fig} the separate transitions can be distinguished. There is a clear pattern with groups of four observed lines (since four different spectra was used for most transitions). The decrease in $\sigma_{\mathrm{Ritz}}-\sigma_{\mathrm{obs}}$ within a group is evident and is a consequence to the way the x-axis is ordered (with increasing lower level and increasing observed wavenumber). It can also be noted that the scatter between the observed wavenumber, measured in different spectra, for the same transition is small and depend mostly on the wavenumber calibration of the different spectra. The final values of the \ion{Cr}{ii} energy levels are presented in  Table~\ref{Cr_levels_table}, together with the uncertainties and the number of observed spectral features used for determining each energy level.

\begin{table}
\begin{minipage}[t]{\columnwidth}
\caption{Low lying \ion{Cr}{ii} energy levels of even parity}
\label{Cr_levels_table}
\centering
\renewcommand{\footnoterule}{}  
\begin{tabular}{l c c r r r}
\hline\hline
\\[-3mm]
Config. & Term & $J$ & Energy & Unc. & Lines\footnote{Number of observed spectral features used in the determination of the energy}\\
 & & & (cm$^{-1}$) & (cm$^{-1}$) & (\#) \\
\hline
\\[-3mm]
3d$^{5}$ & a$^{6}$S & $5/2$ & 0 & 0 & 12\footnote{Energy set to zero by definition} \\
3d$^{4}$($^{5}$D)4s & a$^{6}$D & $1/2$ & 11961.7452 & 0.0012 & 4 \\
 & & $3/2$ & 12032.5433 & 0.0009 & 8   \\
 & & $5/2$ & 12147.7694 & 0.0009 & 10   \\
 & & $7/2$ & 12303.8192 & 0.0008 & 8   \\
 & & $9/2$ & 12496.4549 & 0.0011 & 4   \\
\hline
\end{tabular}
\end{minipage}
\end{table}

\subsection{Ritz wavelengths}

Ritz wavelengths are wavelengths derived from experimentally established energy levels. For \ion{Fe}{ii}, \ion{Ti}{ii} and \ion{Cr}{ii} the energy levels of the lowest even LS terms have been accurately determined in this work. Within and between these terms M1 and E2 transitions can occur following the transition rules in Table~\ref{Rules_table}. For all possible M1 and E2 transitions the vacuum Ritz wavelength was calculated using the energy levels in Table~\ref{Fe_levels_table}, \ref{Ti_levels_table} and \ref{Cr_levels_table}.

For \ion{Fe}{ii} this resulted in a total number of 97 different parity-forbidden transitions within a wavelength range of 0.74--87\,$\mu$m. For most of these transitions the transition probability has been calculated by \citet{Quinet96}, who included only transitions for which $A_{ki}$ was greater than 0.001\,s$^{-1}$. In addition to these \citet{Quinet96} also included transition probabilities for a couple of weaker infrared transitions for comparison with the values by \citet{Nussbaumer88}. The published transition probabilities are the sum of both M1 and E2 transitions (if both contributed) and to be included the different types of radiation for a given transition had to be at least 1\,\% of the sum of M1 and E2 contributions. \citet{Quinet96} published both Superstructure and relativistic Hartree-Fock calculations for comparison, but states that the Superstructure calculations should be more accurate. Therefore, we have chosen to only present these. The new \ion{Fe}{ii} Ritz wavelengths together with the transition probabilities and radiation types from \citet{Quinet96} are presented in Table~\ref{Fe_ritzterm_table}, sorted by multiplet.

For \ion{Ti}{ii} a total number of 24 different parity-forbidden transitions are presented. The wavelength range of the transitions is 8.9--130\,$\mu$m. Unfortunately, to our knowledge, no transition probabilities for these transitions have been published. The new \ion{Ti}{ii} Ritz wavelengths are presented in Table~\ref{Ti_ritzterm_table}, sorted by multiplet.

For \ion{Cr}{ii} a total number of 12 different parity-forbidden transitions, within a wavelength range of 0.80--140\,$\mu$m, are presented. Calculated transition probabilities for the a$^{6}$S-a$^{6}$D transitions have been published by \citet{Quinet97}, using a pseudo relativistic Hartree-Fock method. As for the \ion{Fe}{ii} transitions \citet{Quinet97} only included transitions for which $A_{ki}$ was greater than 0.001\,s$^{-1}$, and the published transition probabilities are the sum of both M1 and E2 transitions (if both contributed). The new \ion{Cr}{ii} Ritz wavelengths together with the transition probabilities and radiation types from \citet{Quinet97} are presented in Table~\ref{Cr_ritzterm_table}, sorted by multiplet.

\subsection{Wavelengths in air}

Deriving the vacuum Ritz wavelengths from the experimentally determined energy levels is straight forward, since
\begin{equation}\label{Lambdavac_Eq}
\lambda_{\mathrm{vac}}=10^{8}\times\frac{1}{\sigma}\,,
\end{equation}
where $\lambda_{\mathrm{vac}}$ is in \AA\ and $\sigma$ is in cm$^{-1}$. The air wavelengths are, on the other hand, depending on the refractive index of air, $n$, so that
\begin{equation}\label{Lambdaair_Eq}
\lambda_{\mathrm{air}}=10^{8}\times\frac{1}{n\sigma}\,.
\end{equation}
The refractive index of air has been investigated at length (see e.g. \citet{Edlen66,Birch93,Ciddor96,Bonsch98}) and is accurately determined for the Visible to UV wavelength range, where the correction from vacuum to air is small. However for the infrared region the corrections become larger and the refractive index is not known to as good accuracy, due to the effects of water vapor and carbon dioxide. \citet{Ciddor96} focused on the visible to near infrared region, but states that the results agreed with the modified Edl\'en equations by \citet{Birch94}. We have chosen to present the vacuum wavelengths together with air wavelengths derived by the \citet{Birch94} formula (based on the formula by \citet{Edlen66}) for dry air (0.045\,\% by volume of carbon dioxide) at $15\degr \mathrm{C}$ and atmospheric pressure, as recommended by the CRC handbook of Chemistry \& Physics \citep{Lide06}. Note however that the accuracy for wavelengths longer than 2\,$\mu$m is decreasing. The refractive index of air is depending on e.g. the temperature, pressure and humidity so for accurate use of air wavelengths the correct value should be calculated from the wavenumbers or the vacuum wavelengths.

\section{Uncertainties}\label{Uncert_sect}

The Ritz wavelengths presented in this work are derived from experimentally determined energy levels, which in turn are determined by the measured wavenumbers of observed spectral lines. The uncertainties in the derived Ritz wavelengths for the M1 and E2 transitions depend thus on the uncertainties in the measured wavenumbers of the observed E1 transitions.

\subsection{Observed wavenumbers}\label{Unc_obs_sect}

The uncertainties in the measured wavenumbers of the observed E1 transitions depend on both the uncertainty of the wavenumber calibration and on the uncertainty of the determination of the line position. The standard deviation $s(\sigma)$ of the line position $\sigma$ of a fitted line profile can be written as \citep{Brault87,Sikstrom02}
\begin{equation}\label{unceq}
s(\sigma)=\alpha_{\sigma}\frac{\sqrt{\mathrm{d}x}}{S}\sqrt{w}=\alpha_{\sigma}\frac{w}{S\sqrt{n}}\,,
\end{equation}
where $S$ is the signal-to-noise ratio for the line, $\mathrm{d}x$ is the resolution interval and $w$ is the FWHM of the line. The number of points across the width of the line is given by $n=w/\mathrm{d}x$. $\alpha_{\sigma}$ is a numerical constant depending on the shape of the line, see~\citet{Sikstrom02}.

The uncertainty from the calibration is derived from the uncertainties of the standard wavenumbers by \citet{Whaling95} and from the uncertainties in the determinations of the line positions of the calibration lines in the spectra. With all possible systematic effects and uncertainties we used a conservative lower limit of 0.0010\,cm$^{-1}$ for the total uncertainties of the absolute wavenumbers of the observed lines, see e.g \citet{Learner88}.

The relative wavenumbers can be determined more accurately, since the uncertainty in the calibration and other systematic uncertainties have less significance, see~\citet{Aldenius06}. This leads to the fact that the energy levels, and Ritz wavelengths, can be more accurately determined than the observed wavenumbers.

\subsection{Energy Levels}

The uncertainties in the fitted energy levels are given by the square root of the covariance in the fitting procedure, see \citet{Oberg07}. In the least-square fitting routine an additional option of modifying the uncertainties of the observed wavenumbers is available. Two factors representing global corrections for underestimated uncertainties are included. This option was not used in this investigation, since the uncertainties of the observed wavenumbers were conservatively estimated and the calculated $\chi ^{2}$ values from the least-square fit were clearly smaller than the degrees of freedom of the system.

The uncertainties of the energy levels range from 0.0006 to 0.0016\,cm$^{-1}$, where the lowest energies generally were determined more accurately due to the larger number of transitions connecting them to the ground state. 

\subsection{Ritz wavelengths}

Ritz wavenumbers were calculated between energy levels included in the fit and the uncertainty for this wavenumber was given by the square root of the variance. The uncertainty of the corresponding vacuum wavelength is related to the uncertainty of the wavenumber as
\begin{equation}
\Delta \lambda_{\mathrm{vac}} = 10^{8}\times \frac{\Delta \sigma}{\sigma ^{2}}\,,
\label{Uncwaveleq}
\end{equation}
and the wavelength uncertainties therefore increase for lower wavenumbers. The uncertainties in the Ritz wavenumbers range from 0.0005 to 0.0017\,cm$^{-1}$, while the uncertainties in the vacuum wavelengths range from 0.00056 to 23\,\AA , depending greatly on the wavelength of the transition.

\section{Results}

For \ion{Fe}{ii} the experimentally determined energy levels are presented in Table~\ref{Fe_levels_table} together with the estimated uncertainties. The energies range from 0 to 13900\,cm$^{-1}$ and include the four lowest terms (a$^{6}$D, a$^{4}$F, a$^{4}$D and a$^{4}$P). The number of observed spectral lines used to determine each energy is also stated. Note that this number includes one or more observed lines of the same transition, since the spectral lines were in general observed in several of the acquired spectra. The derived Ritz wavenumbers, wavelengths and uncertainties for the parity-forbidden lines are presented in Table~\ref{Fe_ritzterm_table}, together with calculated transition probabilities by \cite{Quinet96}. An extensive list containing wavelengths of parity-forbidden \ion{Fe}{ii} transitions has previously been published by \citet{Johansson77},  who used a combination of grating spectroscopy and interferometric measurements. The study included wavelengths between 3000 and 12000\,\AA\ and consisted of 281 parity-forbidden transitions. For lines included in both our work and the work of \citet{Johansson77}, the uncertainties in our wavelengths are almost an order of magnitude lower.

For \ion{Ti}{ii} the energy levels are presented in Table~\ref{Ti_levels_table} together with the estimated uncertainties. The energies range from 0 to 1200\,cm$^{-1}$ and include the two lowest terms (a$^{4}$F and b$^{4}$F). The derived Ritz wavenumbers, wavelengths and uncertainties for the parity-forbidden lines are presented in Table~\ref{Ti_ritzterm_table}. No published experimental or calculated transition probabilities were available for these transitions.

For \ion{Cr}{ii} the energy levels are presented in Table~\ref{Cr_levels_table} together with the estimated uncertainties. The energies range from 0 to 12500\,cm$^{-1}$ and include the two lowest terms (a$^{6}$S and a$^{6}$D). The derived Ritz wavenumbers, wavelengths and uncertainties for the parity-forbidden lines are presented in Table~\ref{Cr_ritzterm_table}, together with calculated transition probabilities by \cite{Quinet97}.

The predicted strongest lines have been collected into a separate finding list arranged by wavelength in Table~\ref{Strong_table}. These transitions have been chosen within the 0.76--15\,$\mu$m region, which covers both the CRIRES region and part of the SPITZER region. \ion{Fe}{ii} and \ion{Cr}{ii} lines, which have calculated transition probabilities larger than 0.002\,s$^{-1}$ \citep{Quinet96,Quinet97} are included. Since no transition probabilities are available for the \ion{Ti}{ii} transitions we have chosen to include the a$^{4}$F--b$^{4}$F transitions with $\Delta J=0,\pm 1$ in Table~\ref{Strong_table}.

\onllongtab{6}{
{\small
\begin{longtable}{l l r r r r r @{.} l l r @{.} l c c}
\caption{\label{Fe_ritzterm_table}Infrared \ion{Fe}{ii} Ritz wavelengths of parity-forbidden transitions. Sorted by multiplet}\\
\hline\hline
\\[-3mm]
\multicolumn{2}{c}{Transition} & \multicolumn{2}{c}{Energy (cm$^{-1}$)\,$^{a}$} & Wavenumber & Unc. & \multicolumn{2}{c}{$\lambda_{\mathrm{vac}}$} & Unc. & \multicolumn{2}{c}{$\lambda_{\mathrm{air}}$\,$^{b}$} & $A_{ik}$\,$^{c}$ & Type\\
Lower & Upper & Lower & Upper & (cm$^{-1}$) & (cm$^{-1}$) & \multicolumn{2}{c}{(\AA )} & (\AA ) & \multicolumn{2}{c}{(\AA )} & (s$^{-1}$) & \\
\hline
\endfirsthead
\caption{continued.}\\
\hline\hline
\\[-3mm]
\multicolumn{2}{c}{Transition} & \multicolumn{2}{c}{Energy (cm$^{-1}$)\,$^{a}$} & Wavenumber & Unc. & \multicolumn{2}{c}{$\lambda_{\mathrm{vac}}$} & Unc. & \multicolumn{2}{c}{$\lambda_{\mathrm{air}}$\,$^{b}$} & $A_{ik}$\,$^{c}$ & Type\\
Lower & Upper & Lower & Upper & (cm$^{-1}$) & (cm$^{-1}$) & \multicolumn{2}{c}{(\AA )} & (\AA ) & \multicolumn{2}{c}{(\AA )} & (s$^{-1}$) & \\
\hline
\endhead
\hline
\endfoot
\\[-3mm]
a$^{6}$D$_{9/2}$ & a$^{6}$D$_{7/2}$ & 0   & 384   & 384.7868   &  0.0007 & 259884&18   & 0.47   & 259813&35  & 2.13$\times10^{-3}$ & M1   \\
a$^{6}$D$_{9/2}$ & a$^{6}$D$_{5/2}$ & 0   & 667   & 667.6830   &  0.0008 & 149771&66   & 0.17   & 149730&84  &    \\
a$^{6}$D$_{7/2}$ & a$^{6}$D$_{5/2}$ & 384 & 667   & 282.8963   &  0.0005 & 353486&48   & 0.66   & 353390&13  & 1.57$\times10^{-3}$ & M1   \\
a$^{6}$D$_{7/2}$ & a$^{6}$D$_{3/2}$ & 384 & 862   & 477.8240   &  0.0007 & 209282&08   & 0.31   & 209225&04  &    \\
a$^{6}$D$_{5/2}$ & a$^{6}$D$_{3/2}$ & 667 & 862   & 194.9277   &  0.0006 & 513010&6    & 1.6    & 512870&8   & 7.18$\times10^{-4}$ & M1   \\
a$^{6}$D$_{5/2}$ & a$^{6}$D$_{1/2}$ & 667 & 977   & 309.3658   &  0.0007 & 323241&89   & 0.72   & 323153&79  &    \\ \vspace{2mm}
a$^{6}$D$_{3/2}$ & a$^{6}$D$_{1/2}$ & 862 & 977   & 114.4381   &  0.0006 & 873834&8    & 4.9    & 873596&7   & 1.89$\times10^{-4}$ & M1   \\

a$^{6}$D$_{9/2}$ & a$^{4}$F$_{9/2}$ & 0   & 1872  & 1872.6005  & 0.0013 &  53401&674   & 0.036    & 53387&117  & 9.15$\times10^{-5}$ & M1   \\
a$^{6}$D$_{9/2}$ & a$^{4}$F$_{7/2}$ & 0   & 2430  & 2430.1369  & 0.0013 &  41149&946   & 0.021    & 41138&726  & 3.04$\times10^{-5}$ & M1   \\
a$^{6}$D$_{9/2}$ & a$^{4}$F$_{5/2}$ & 0   & 2837  & 2837.9803  & 0.0014 &  35236&326   & 0.017    & 35226&718  &    \\
a$^{6}$D$_{7/2}$ & a$^{4}$F$_{9/2}$ & 384 & 1872  & 1487.8137	& 0.0012 &  67212&717  & 0.054  & 67194&396  & 8.36$\times10^{-6}$ & M1   \\
a$^{6}$D$_{7/2}$ & a$^{4}$F$_{7/2}$ & 384 & 2430  & 2045.3501	& 0.0012 &  48891&386  & 0.029  & 48878&057  & 6.39$\times10^{-5}$ & M1   \\
a$^{6}$D$_{7/2}$ & a$^{4}$F$_{5/2}$ & 384 & 2837  & 2453.1935	& 0.0013 &  40763&193  & 0.021  & 40752&079  & 2.31$\times10^{-5}$ & M1   \\ 
a$^{6}$D$_{7/2}$ & a$^{4}$F$_{3/2}$ & 384 & 3117  & 2732.7020	& 0.0014 &  36593&818  & 0.019  & 36583&840  &    \\
a$^{6}$D$_{5/2}$ & a$^{4}$F$_{9/2}$ & 667 & 1872  & 1204.9174  &  0.0012 &  82993&239  & 0.085  & 82970&617  &    \\
a$^{6}$D$_{5/2}$ & a$^{4}$F$_{7/2}$ & 667 & 2430  & 1762.4538  &  0.0012 &  56739&075  & 0.040  & 56723&608  & 1.14$\times10^{-5}$ & M1   \\
a$^{6}$D$_{5/2}$ & a$^{4}$F$_{5/2}$ & 667 & 2837  & 2170.2973  &  0.0013 &  46076&637  & 0.028  & 46064&076  & 3.69$\times10^{-5}$ & M1   \\
a$^{6}$D$_{5/2}$ & a$^{4}$F$_{3/2}$ & 667 & 3117  & 2449.8058  &  0.0015 &  40819&563  & 0.024  & 40808&434  & 9.87$\times10^{-6}$ & M1   \\
a$^{6}$D$_{3/2}$ & a$^{4}$F$_{7/2}$ & 862 & 2430  & 1567.5261  &  0.0013 &  63794&792  & 0.054  & 63777&402  &    \\
a$^{6}$D$_{3/2}$ & a$^{4}$F$_{5/2}$ & 862 & 2837  & 1975.3695  &  0.0014 &  50623&440  & 0.036  & 50609&639  & 8.89$\times10^{-6}$ & M1   \\
a$^{6}$D$_{3/2}$ & a$^{4}$F$_{3/2}$ & 862 & 3117  & 2254.8780  &  0.0015 &  44348&297  & 0.030  & 44336&206  & 1.51$\times10^{-5}$ & M1   \\
a$^{6}$D$_{1/2}$ & a$^{4}$F$_{5/2}$ & 977 & 2837  & 1860.9314  &  0.0015 & 53736&531   & 0.042  & 53721&883  &    \\ \vspace{2mm}
a$^{6}$D$_{1/2}$ & a$^{4}$F$_{3/2}$ & 977 & 3117  & 2140.4399  &  0.0016 & 46719&368   & 0.035  & 46706&631  & 3.74$\times10^{-6}$ & M1   \\

a$^{6}$D$_{9/2}$ & a$^{4}$D$_{7/2}$ & 0  & 7955  & 7955.3189  & 0.0012 &  12570&2064  & 0.0019  & 12566&7681 & 4.74$\times10^{-3}$ & M1   \\
a$^{6}$D$_{9/2}$ & a$^{4}$D$_{5/2}$ & 0  & 8391  & 8391.9562  & 0.0012 &  11916&1728  & 0.0018  & 11912&9121 & 5.84$\times10^{-6}$ & E2   \\
a$^{6}$D$_{7/2}$ & a$^{4}$D$_{7/2}$ & 384 & 7955  & 7570.5321	& 0.0011 &  13209&1112 & 0.0020 & 13205&4994 & 1.31$\times10^{-3}$ & M1   \\
a$^{6}$D$_{7/2}$ & a$^{4}$D$_{5/2}$ & 384 & 8391  & 8007.1694	& 0.0012 &  12488&8079 & 0.0018 & 12485&3917 & 3.78$\times10^{-4}$ & M1   \\
a$^{6}$D$_{7/2}$ & a$^{4}$D$_{3/2}$ & 384 & 8680  & 8295.6840	& 0.0012 &  12054&4611 & 0.0018 & 12051&1629 & 2.62$\times10^{-6}$ & E2   \\
a$^{6}$D$_{5/2}$ & a$^{4}$D$_{7/2}$ & 667 & 7955  & 7287.6358  &  0.0012 &  13721&8712 & 0.0022 & 13718&1201 & 8.42$\times10^{-4}$ & M1   \\
a$^{6}$D$_{5/2}$ & a$^{4}$D$_{5/2}$ & 667 & 8391  & 7724.2731  &  0.0012 &  12946&2020 & 0.0020 & 12942&6616 & 1.98$\times10^{-3}$ & M1   \\
a$^{6}$D$_{5/2}$ & a$^{4}$D$_{3/2}$ & 667 & 8680  & 8012.7877  &  0.0013 &  12480&0511 & 0.0020 & 12476&6373 & 4.64$\times10^{-5}$ & M1   \\
a$^{6}$D$_{5/2}$ & a$^{4}$D$_{1/2}$ & 667 & 8846  & 8179.0992  &  0.0015 &  12226&2853 & 0.0022 & 12222&9404 &    \\
a$^{6}$D$_{3/2}$ & a$^{4}$D$_{7/2}$ & 862 & 7955  & 7092.7081  &  0.0013 &  14098&9871 & 0.0025 & 14095&1335 &    \\
a$^{6}$D$_{3/2}$ & a$^{4}$D$_{5/2}$ & 862 & 8391  & 7529.3454  &  0.0013 &  13281&3671 & 0.0023 & 13277&7357 & 1.17$\times10^{-3}$ & M1   \\
a$^{6}$D$_{3/2}$ & a$^{4}$D$_{3/2}$ & 862 & 8680  & 7817.8600  &  0.0014 &  12791&2242 & 0.0022 & 12787&7260 & 2.45$\times10^{-3}$ & M1   \\
a$^{6}$D$_{3/2}$ & a$^{4}$D$_{1/2}$ & 862 & 8846  & 7984.1714  &  0.0016 &  12524&7812 & 0.0024 & 12521&3552 & 6.46$\times10^{-4}$ & M1   \\
a$^{6}$D$_{1/2}$ & a$^{4}$D$_{5/2}$ & 977 & 8391  & 7414.9073  &  0.0014 & 13486&3453  & 0.0025 & 13482&6582 &    \\
a$^{6}$D$_{1/2}$ & a$^{4}$D$_{3/2}$ & 977 & 8680  & 7703.4219  &  0.0014 & 12981&2441  & 0.0024 & 12977&6942 & 1.08$\times10^{-3}$ & M1   \\ \vspace{2mm}
a$^{6}$D$_{1/2}$ & a$^{4}$D$_{1/2}$ & 977 & 8846  & 7869.7333  &  0.0016 & 12706&9109  & 0.0026 & 12703&4355 & 3.32$\times10^{-3}$ & M1   \\

a$^{6}$D$_{9/2}$ & a$^{4}$P$_{5/2}$ & 0   & 13474 & 13474.4487 &  0.0013 &   7421&45389 & 0.00072 &  7419&41022 &    \\
a$^{6}$D$_{7/2}$ & a$^{4}$P$_{5/2}$ & 384 & 13474 & 13089.6619 &  0.0012 &   7639&61670 & 0.00073 &  7637&51419 & 6.64$\times10^{-3}$ & M1   \\
a$^{6}$D$_{7/2}$ & a$^{4}$P$_{3/2}$ & 384 & 13673 & 13288.4152 &  0.0014 &   7525&35188 & 0.00077 &  7523&28019 & 9.25$\times10^{-5}$ & E2   \\
a$^{6}$D$_{5/2}$ & a$^{4}$P$_{5/2}$ & 667 & 13474 & 12806.7657 &  0.0013 &   7808&37273 & 0.00078 &  7806&22468 & 1.20$\times10^{-4}$ & M1,E2   \\
a$^{6}$D$_{5/2}$ & a$^{4}$P$_{3/2}$ & 667 & 13673 & 13005.5189 &  0.0014 &   7689&04345 & 0.00083 &  7686&92760 & 6.81$\times10^{-3}$ & M1   \\
a$^{6}$D$_{5/2}$ & a$^{4}$P$_{1/2}$ & 667 & 13904 & 13237.1768 &  0.0016 &   7554&48095 & 0.00092 &  7552&40140 & 2.03$\times10^{-4}$ & E2   \\
a$^{6}$D$_{3/2}$ & a$^{4}$P$_{5/2}$ & 862 & 13474 & 12611.8379 &  0.0014 &   7929&05843 & 0.00086 &  7926&87779 & 5.02$\times10^{-4}$ & M1   \\
a$^{6}$D$_{3/2}$ & a$^{4}$P$_{3/2}$ & 862 & 13673 & 12810.5912 &  0.0015 &   7806&04101 & 0.00090 &  7803&89359 & 3.76$\times10^{-4}$ & M1,E2   \\
a$^{6}$D$_{3/2}$ & a$^{4}$P$_{1/2}$ & 862 & 13904 & 13042.2491 &  0.0017 &   7667&38923 & 0.00099 &  7665&27922 & 6.23$\times10^{-3}$ & M1   \\
a$^{6}$D$_{1/2}$ & a$^{4}$P$_{5/2}$ & 977 & 13474 & 12497.3998 &  0.0014 &   8001&66445 & 0.00091 &  7999&46420 & 6.69$\times10^{-6}$ & E2   \\
a$^{6}$D$_{1/2}$ & a$^{4}$P$_{3/2}$ & 977 & 13673 & 12696.1531 &  0.0015 &   7876&40158 & 0.00095 &  7874&23516 & 9.80$\times10^{-4}$ & M1   \\ \vspace{1mm}
a$^{6}$D$_{1/2}$ & a$^{4}$P$_{1/2}$ & 977 & 13904 & 12927.8110 &  0.0017 &   7735&2616  & 0.0010  &  7733&1333  & 1.93$\times10^{-3}$ & M1   \\
\newpage
\\[-3mm]
a$^{4}$F$_{9/2}$ & a$^{4}$F$_{7/2}$ & 1872 & 2430  & 557.5364   & 0.0008 & 179360&49   & 0.27   & 179311&60  & 5.84$\times10^{-3}$ & M1   \\
a$^{4}$F$_{9/2}$ & a$^{4}$F$_{5/2}$ & 1872 & 2837  & 965.3798   & 0.0010 & 103586&17   & 0.11   & 103557&94  &    \\
a$^{4}$F$_{7/2}$ & a$^{4}$F$_{5/2}$ & 2430 & 2837  & 407.8434	& 0.0009 & 245192&12   & 0.52   & 245125&29  & 3.92$\times10^{-3}$ & M1   \\
a$^{4}$F$_{7/2}$ & a$^{4}$F$_{3/2}$ & 2430 & 3117  & 687.3519	& 0.0011 & 145485&88   & 0.23   & 145446&22  &    \\
\vspace{2mm}
a$^{4}$F$_{5/2}$ & a$^{4}$F$_{3/2}$ & 2837 & 3117  & 279.5085	& 0.0010 & 357770&9    & 1.2    & 357673&4   & 1.41$\times10^{-3}$ & M1   \\

a$^{4}$F$_{9/2}$ & a$^{4}$D$_{7/2}$ & 1872 & 7955  & 6082.7184  & 0.0007 &  16440&0180 & 0.0020 & 16435&5279 & 5.98$\times10^{-3}$ & E2   \\
a$^{4}$F$_{9/2}$ & a$^{4}$D$_{5/2}$ & 1872 & 8391  & 6519.3557 	& 0.0008 &  15338&9391 & 0.0019 & 15334&7484 & 3.12$\times10^{-3}$ & E2   \\
a$^{4}$F$_{7/2}$ & a$^{4}$D$_{7/2}$ & 2430 & 7955  & 5525.1820	& 0.0007 &  18098&9513 & 0.0023 & 18094&0099 & 1.32$\times10^{-3}$ & E2   \\
a$^{4}$F$_{7/2}$ & a$^{4}$D$_{5/2}$ & 2430 & 8391  & 5961.8193	& 0.0007 &  16773&4034 & 0.0020 & 16768&8226 & 2.49$\times10^{-3}$ & E2   \\
a$^{4}$F$_{7/2}$ & a$^{4}$D$_{3/2}$ & 2430 & 8680  & 6250.3339	& 0.0008 &  15999&1453 & 0.0020 & 15994&7751 & 4.18$\times10^{-3}$ & E2   \\
a$^{4}$F$_{5/2}$ & a$^{4}$D$_{7/2}$ & 2837 & 7955  & 5117.3386	& 0.0009 &  19541&4079 & 0.0033 & 19536&0740 & 1.46$\times10^{-4}$ & E2   \\
a$^{4}$F$_{5/2}$ & a$^{4}$D$_{5/2}$ & 2837 & 8391  & 5553.9759	& 0.0008 &  18005&1197 & 0.0025 & 18000&2038 & 1.82$\times10^{-3}$ & E2   \\
a$^{4}$F$_{5/2}$ & a$^{4}$D$_{3/2}$ & 2837 & 8680  & 5842.4904	& 0.0007 &  17115&9886 & 0.0022 & 17111&3147 & 1.18$\times10^{-3}$ & E2   \\
a$^{4}$F$_{5/2}$ & a$^{4}$D$_{1/2}$ & 2837 & 8846  & 6008.8019	& 0.0010 &  16642&2527 & 0.0027 & 16637&7076 & 4.75$\times10^{-3}$ & E2   \\
a$^{4}$F$_{3/2}$ & a$^{4}$D$_{7/2}$ & 3117 & 7955  & 4837.8301	& 0.0011 & 20670&4243  & 0.0045 & 20664&7831 &    \\
a$^{4}$F$_{3/2}$ & a$^{4}$D$_{5/2}$ & 3117 & 8391  & 5274.4674	& 0.0010 & 18959&2603  & 0.0034 & 18954&0849 & 2.98$\times10^{-4}$ & E2   \\
a$^{4}$F$_{3/2}$ & a$^{4}$D$_{3/2}$ & 3117 & 8680  & 5562.9820	& 0.0009 & 17975&9706  & 0.0028 & 17971&0627 & 2.12$\times10^{-3}$ & E2   \\ \vspace{2mm}
a$^{4}$F$_{3/2}$ & a$^{4}$D$_{1/2}$ & 3117 & 8846  & 5729.2934	& 0.0010 & 17454&1593  & 0.0030 & 17449&3933 & 2.47$\times10^{-3}$ & E2   \\

a$^{4}$F$_{9/2}$ & a$^{4}$P$_{5/2}$ & 1872 & 13474 & 11601.8483 & 0.0009 &  8619&31631 & 0.00067 & 8616&94915 & 3.56$\times10^{-2}$ & E2   \\
a$^{4}$F$_{7/2}$ & a$^{4}$P$_{5/2}$ & 2430 & 13474 & 11044.3119	& 0.0008 &  9054&43465 & 0.00068 & 9051&94977 & 8.83$\times10^{-3}$ & M1,E2   \\
a$^{4}$F$_{7/2}$ & a$^{4}$P$_{3/2}$ & 2430 & 13673 & 11243.0651	& 0.0010 &  8894&37171 & 0.00076 & 8891&93015 & 2.21$\times10^{-2}$ & E2   \\
a$^{4}$F$_{5/2}$ & a$^{4}$P$_{5/2}$ & 2837 & 13474 & 10636.4684	& 0.0009 &  9401&61678 & 0.00082 & 9399&03793 & 1.68$\times10^{-3}$ & M1,E2   \\
a$^{4}$F$_{5/2}$ & a$^{4}$P$_{3/2}$ & 2837 & 13673 & 10835.2216	& 0.0010 &  9229&16055 & 0.00088 & 9226&62838 & 1.29$\times10^{-2}$ & E2   \\
a$^{4}$F$_{5/2}$ & a$^{4}$P$_{1/2}$ & 2837 & 13904 & 11066.8795	& 0.0013 &  9035&97077 & 0.0010  & 9033&4909  & 1.61$\times10^{-2}$ & E2   \\
a$^{4}$F$_{3/2}$ & a$^{4}$P$_{5/2}$ & 3117 & 13474 & 10356.9599	& 0.0011 &  9655&3429  & 0.0010  & 9652&6954  & 1.37$\times10^{-4}$ & E2   \\
a$^{4}$F$_{3/2}$ & a$^{4}$P$_{3/2}$ & 3117 & 13673 & 10555.7131	& 0.0012 &  9473&5428  & 0.0011  & 9470&9444  & 3.65$\times10^{-3}$ & E2   \\ \vspace{2mm}
a$^{4}$F$_{3/2}$ & a$^{4}$P$_{1/2}$ & 3117 & 13904 & 10787.3710	& 0.0014 &  9270&0992  & 0.0012  & 9267&5560  & 2.13$\times10^{-2}$ & E2   \\

a$^{4}$D$_{7/2}$ & a$^{4}$D$_{5/2}$ & 7955 & 8391  & 436.6373	& 0.0007 & 229023&03   & 0.35   & 228960&61  & 2.56$\times10^{-3}$ & M1   \\
a$^{4}$D$_{7/2}$ & a$^{4}$D$_{3/2}$ & 7955 & 8680  & 725.1519	& 0.0008 & 137902&14   & 0.15   & 137864&56  &    \\
a$^{4}$D$_{5/2}$ & a$^{4}$D$_{3/2}$ & 8391 & 8680  & 288.5146   & 0.0007 & 346602&93   & 0.79   & 346508&46  & 1.36$\times10^{-3}$ & M1   \\
a$^{4}$D$_{5/2}$ & a$^{4}$D$_{1/2}$ & 8391 & 8846  & 454.8261	& 0.0010 & 219864&27   & 0.48   & 219804&35  &    \\ \vspace{2mm}
a$^{4}$D$_{3/2}$ & a$^{4}$D$_{1/2}$ & 8680 & 8846  & 166.3115	& 0.0009 & 601281&4    & 3.2    & 601117&6   & 3.71$\times10^{-4}$ & M1   \\ 

a$^{4}$D$_{7/2}$ & a$^{4}$P$_{5/2}$ & 7955 & 13474 & 5519.1299	& 0.0007 &  18118&7982 & 0.0024 & 18113&8514 & 2.23$\times10^{-3}$ & M1,E2   \\
a$^{4}$D$_{7/2}$ & a$^{4}$P$_{3/2}$ & 7955 & 13673 & 5717.8831	& 0.0009 &  17488&9900 & 0.0028 & 17484&2146 & 2.25$\times10^{-3}$ & E2   \\
a$^{4}$D$_{5/2}$ & a$^{4}$P$_{5/2}$ & 8391 & 13474 & 5082.4926	& 0.0007 &  19675&3854 & 0.0028 & 19670&0151 & 1.06$\times10^{-3}$ & M1,E2   \\
a$^{4}$D$_{5/2}$ & a$^{4}$P$_{3/2}$ & 8391 & 13673 & 5281.2458	& 0.0009 &  18934&9264 & 0.0032 & 18929&7575 & 6.28$\times10^{-5}$ & M1,E2   \\
a$^{4}$D$_{5/2}$ & a$^{4}$P$_{1/2}$ & 8391 & 13904 & 5512.9037	& 0.0012 &  18139&2612 & 0.0038 & 18134&3089 & 2.76$\times10^{-3}$ & E2   \\
a$^{4}$D$_{3/2}$ & a$^{4}$P$_{5/2}$ & 8680 & 13474 & 4793.9780	& 0.0008 &  20859&5034 & 0.0036 & 20853&8108 & 3.67$\times10^{-4}$ & M1,E2   \\
a$^{4}$D$_{3/2}$ & a$^{4}$P$_{3/2}$ & 8680 & 13673 & 4992.7312	& 0.0010 &  20029&1176 & 0.0039 & 20023&6509 & 6.35$\times10^{-4}$ & M1,E2   \\ 
a$^{4}$D$_{3/2}$ & a$^{4}$P$_{1/2}$ & 8680 & 13904 & 5224.3891	& 0.0012 &  19140&9940 & 0.0043 & 19135&7690 & 9.09$\times10^{-4}$ & E2   \\
a$^{4}$D$_{1/2}$ & a$^{4}$P$_{5/2}$ & 8846 & 13474 & 4627.6665  & 0.0011 &  21609&1630 & 0.0052 & 21603&2663 & 3.36$\times10^{-5}$ & E2   \\ 
a$^{4}$D$_{1/2}$ & a$^{4}$P$_{3/2}$ & 8846 & 13673 & 4826.4197  & 0.0012 &  20719&2921 & 0.0053 & 20713&6376 & 5.77$\times10^{-4}$ & M1,E2   \\ \vspace{2mm}
a$^{4}$D$_{1/2}$ & a$^{4}$P$_{1/2}$ & 8846 & 13904 & 5058.0776  & 0.0014 &  19770&3569 & 0.0054 & 19764&9607  & 4.04$\times10^{-5}$ & M1   \\

a$^{4}$P$_{5/2}$ & a$^{4}$P$_{3/2}$ & 13474 & 13673 & 198.7532 & 0.0010 & 503136&5     & 2.4    & 502999&4   & 1.86$\times10^{-4}$ & M1   \\
a$^{4}$P$_{5/2}$ & a$^{4}$P$_{1/2}$ & 13474 & 13904 & 430.4111 & 0.0012 & 232336&01    & 0.67   & 232272&68  &    \\ \vspace{1mm}
a$^{4}$P$_{3/2}$ & a$^{4}$P$_{1/2}$ & 13673 & 13904 & 231.6579 & 0.0014	& 431671&0     & 2.6    & 431553&4   & 5.51$\times10^{-4}$ & M1   \\
\end{longtable}

{\footnotesize
\indent$^{a}$ Truncated energy. For more exact value see Table~\ref{Fe_levels_table}\\
\indent $^{b}$ $\lambda_{\mathrm{air}}$ is calculated from the wavenumber using the modified \citet{Edlen66} dispersion formula by \citet{Birch94} for standard air\\
\indent $^{c}$ Transition probabilities taken from the Superstructure calculations by \citet{Quinet96} 
}
}
}

\onllongtab{7}{
{\small
\begin{longtable}{l l r r r r r @{.} l l r @{.} l}
\caption{\label{Ti_ritzterm_table}Infrared \ion{Ti}{ii} Ritz wavelengths of parity-forbidden transitions. Sorted by multiplet}\\
\hline\hline
\\[-3mm]
\multicolumn{2}{c}{Transition} & \multicolumn{2}{c}{Energy (cm$^{-1}$)$^{a}$} & Wavenumber & Unc. & \multicolumn{2}{c}{$\lambda_{\mathrm{vac}}$} & Unc. & \multicolumn{2}{c}{$\lambda_{\mathrm{air}}$\,$^{b}$} \\
Lower & Upper & Lower & Upper & (cm$^{-1}$) & (cm$^{-1}$) & \multicolumn{2}{c}{(\AA )} & (\AA ) & \multicolumn{2}{c}{(\AA )} \\
\hline
\endfirsthead
\caption{continued.}\\
\hline\hline
\\[-3mm]
\multicolumn{2}{c}{Transition} & \multicolumn{2}{c}{Energy (cm$^{-1}$)$^{a}$} & Wavenumber & Unc. & \multicolumn{2}{c}{$\lambda_{\mathrm{vac}}$} & Unc. & \multicolumn{2}{c}{$\lambda_{\mathrm{air}}$\,$^{b}$} \\
Lower & Upper & Lower & Upper & (cm$^{-1}$) & (cm$^{-1}$) & \multicolumn{2}{c}{(\AA )} & (\AA ) & \multicolumn{2}{c}{(\AA )} \\
\hline
\endhead
\hline
\endfoot
\\[-3mm]
a$^{4}$F$_{3/2}$ & a$^{4}$F$_{5/2}$ & 0    & 94   & 94.1105   & 0.0006 & 1062580&5   &   6.3 & 1062290&9     \\
a$^{4}$F$_{3/2}$ & a$^{4}$F$_{7/2}$ & 0    & 225  & 225.7014  & 0.0007 &  443063&3   &	 1.3 &  442942&5     \\                          
a$^{4}$F$_{5/2}$ & a$^{4}$F$_{7/2}$ & 94   & 225  & 131.5909  & 0.0005 &  759931&1   &   3.0 &  759724&0     \\
a$^{4}$F$_{5/2}$ & a$^{4}$F$_{9/2}$ & 94   & 393  & 299.3337  & 0.0007 &  334075&32  &  0.76 &  333984&27    \\ \vspace{2mm}
a$^{4}$F$_{7/2}$ & a$^{4}$F$_{9/2}$ & 225  & 393  & 167.7428  & 0.0006 &  596150&7   &   2.1 &  595988&2     \\

a$^{4}$F$_{3/2}$ & b$^{4}$F$_{3/2}$ & 0    & 907  & 907.9646  & 0.0008 &  110136&46  &  0.10 & 110106&44     \\
a$^{4}$F$_{3/2}$ & b$^{4}$F$_{5/2}$ & 0    & 983  & 983.9127  & 0.0006 &  101635&038 & 0.066 & 101607&335    \\
a$^{4}$F$_{3/2}$ & b$^{4}$F$_{7/2}$ & 0    & 1087 & 1087.3530 &	0.0007 &   91966&455 & 0.063 &  91941&388    \\
a$^{4}$F$_{5/2}$ & b$^{4}$F$_{3/2}$ & 94   & 907  & 813.8540  & 0.0008 &  122872&15  &  0.12 & 122838&66     \\
a$^{4}$F$_{5/2}$ & b$^{4}$F$_{5/2}$ & 94   & 983  & 889.8021  & 0.0005 &  112384&535 & 0.066 & 112353&903    \\
a$^{4}$F$_{5/2}$ & b$^{4}$F$_{7/2}$ & 94   & 1087 & 993.2425  & 0.0006 &  100680&350 & 0.061 & 100652&908    \\
a$^{4}$F$_{5/2}$ & b$^{4}$F$_{9/2}$ & 94   & 1215 & 1121.7199 & 0.0008 &   89148&817 & 0.062 &  89124&518    \\
a$^{4}$F$_{7/2}$ & b$^{4}$F$_{3/2}$ & 225  & 907  & 682.2632  & 0.0009 &  146571&01  & 0.19  & 146531&06     \\
a$^{4}$F$_{7/2}$ & b$^{4}$F$_{5/2}$ & 225  & 983  & 758.2113  & 0.0006 &  131889&36  & 0.10  & 131853&41     \\
a$^{4}$F$_{7/2}$ & b$^{4}$F$_{7/2}$ & 225  & 1087 & 861.6516  & 0.0005 &  116056&188 & 0.074 & 116024&555    \\
a$^{4}$F$_{7/2}$ & b$^{4}$F$_{9/2}$ & 225  & 1215 & 990.1290  & 0.0007 &  100996&941 & 0.072 & 100969&413    \\
a$^{4}$F$_{9/2}$ & b$^{4}$F$_{5/2}$ & 393  & 983  & 590.4684  & 0.0007 &  169357&06  & 0.21  & 169310&90     \\
a$^{4}$F$_{9/2}$ & b$^{4}$F$_{7/2}$ & 393  & 1087 & 693.9088  & 0.0006 &  144111&16  & 0.13  & 144071&88     \\ \vspace{2mm}
a$^{4}$F$_{9/2}$ & b$^{4}$F$_{9/2}$ & 393  & 1215 & 822.3862  & 0.0006 &  121597&375 & 0.087 & 121564&232    \\

b$^{4}$F$_{3/2}$ & b$^{4}$F$_{5/2}$ & 907  & 983  &  75.9481  & 0.0009 & 1316689&    & 15.   & 1316330&      \\
b$^{4}$F$_{3/2}$ & b$^{4}$F$_{7/2}$ & 907  & 1087 & 179.3884  & 0.0009 &  557449&6   & 2.9   & 557297&6      \\
b$^{4}$F$_{5/2}$ & b$^{4}$F$_{7/2}$ & 983  & 1087 & 103.4403  & 0.0007 &  966740&9   & 6.2   & 966477&4      \\
b$^{4}$F$_{5/2}$ & b$^{4}$F$_{9/2}$ & 983  & 1215 & 231.9177  & 0.0008 &  431187&4   & 1.5   & 431069&9      \\ \vspace{1mm}
b$^{4}$F$_{7/2}$ & b$^{4}$F$_{9/2}$ & 1087 & 1215 & 128.4774  & 0.0008 &  778347&0   & 4.6   & 778134&9      \\
\end{longtable}

{\footnotesize
$^{a}$ Truncated energy. For more exact value see Table~\ref{Ti_levels_table}\\
\indent $^{b}$ $\lambda_{\mathrm{air}}$ is calculated from the wavenumber using the modified \citet{Edlen66} dispersion formula by \citet{Birch94} for standard air
}
}
}

\onllongtab{8}{
{\small
\begin{longtable}{l l r r r r r @{.} l l r @{.} l c c}
\caption{\label{Cr_ritzterm_table}Infrared \ion{Cr}{ii} Ritz wavelengths of parity-forbidden transitions. Sorted by multiplet}\\
\hline\hline
\\[-3mm]
\multicolumn{2}{c}{Transition} & \multicolumn{2}{c}{Energy (cm$^{-1}$)$^{a}$} & Wavenumber & Unc. & \multicolumn{2}{c}{$\lambda_{\mathrm{vac}}$} & Unc. & \multicolumn{2}{c}{$\lambda_{\mathrm{air}}$\,$^{b}$} & $A_{ik}$\,$^{c}$ & Type\\
Lower & Upper & Lower & Upper & (cm$^{-1}$) & (cm$^{-1}$) & \multicolumn{2}{c}{(\AA )} & (\AA ) & \multicolumn{2}{c}{(\AA )} & (s$^{-1}$) & \\
\hline
\endfirsthead
\caption{continued.}\\
\hline\hline
\\[-3mm]
\multicolumn{2}{c}{Transition} & \multicolumn{2}{c}{Energy (cm$^{-1}$)$^{a}$} & Wavenumber & Unc. & \multicolumn{2}{c}{$\lambda_{\mathrm{vac}}$} & Unc. & \multicolumn{2}{c}{$\lambda_{\mathrm{air}}$\,$^{b}$} & $A_{ik}$\,$^{c}$ & Type\\
Lower & Upper & Lower & Upper & (cm$^{-1}$) & (cm$^{-1}$) & \multicolumn{2}{c}{(\AA )} & (\AA ) & \multicolumn{2}{c}{(\AA )} & (s$^{-1}$) & \\
\hline
\endhead
\hline
\endfoot
\\[-3mm]
a$^{6}$S$_{5/2}$ & a$^{6}$D$_{1/2}$ & 0    & 11961   & 11961.7452 & 0.0012 & 8359&98415 & 0.00083 & 8357&68710  & 8.40$\times10^{-2}$ & E2  \\
a$^{6}$S$_{5/2}$ & a$^{6}$D$_{3/2}$ & 0    & 12032   & 12032.5433 & 0.0009 & 8310&79498 & 0.00065 & 8308&51122  & 8.65$\times10^{-2}$ & E2  \\
a$^{6}$S$_{5/2}$ & a$^{6}$D$_{5/2}$ & 0    & 12147   & 12147.7694 & 0.0009 & 8231&96395 & 0.00058 & 8229&70150  & 9.10$\times10^{-2}$ & E2  \\ 
a$^{6}$S$_{5/2}$ & a$^{6}$D$_{7/2}$ & 0    & 12303   & 12303.8192 & 0.0008 & 8127&55768 & 0.00056 & 8125&32343  & 9.76$\times10^{-2}$ & E2  \\ \vspace{2mm}
a$^{6}$S$_{5/2}$ & a$^{6}$D$_{9/2}$ & 0    & 12496   & 12496.4549 & 0.0011 & 8002&26949 & 0.00071 & 8000&06908  & 1.06$\times10^{-1}$ & E2  \\

a$^{6}$D$_{1/2}$ & a$^{6}$D$_{3/2}$ & 11961 & 12032  & 70.7981    & 0.0011 & 1412467&   &   23.   & 1412083&    &   \\
a$^{6}$D$_{1/2}$ & a$^{6}$D$_{5/2}$ & 11961 & 12147  & 186.0243   & 0.0011 & 537564&3   &   3.3   &  537417&8   &   \\
a$^{6}$D$_{3/2}$ & a$^{6}$D$_{5/2}$ & 12032 & 12147  & 115.2262   & 0.0009 & 867858&4   &   6.6   &  867621&8   &   \\
a$^{6}$D$_{3/2}$ & a$^{6}$D$_{7/2}$ & 12032 & 12303  & 271.2759   & 0.0010 & 368628&4   &   1.4   &  368527&9   &   \\ 
a$^{6}$D$_{5/2}$ & a$^{6}$D$_{7/2}$ & 12147 & 12303  & 156.0497   & 0.0009 & 640821&3   &   3.8   &  640646&6   &   \\
a$^{6}$D$_{5/2}$ & a$^{6}$D$_{9/2}$ & 12147 & 12496  & 348.6855   & 0.0012 & 286791&4   &   1.0   &  286713&2   &   \\ \vspace{1mm}
a$^{6}$D$_{7/2}$ & a$^{6}$D$_{9/2}$ & 12303 & 12496  & 192.6358   & 0.0011 & 519114&4   &   2.9   &  518972&9   &   \\
\end{longtable}

{\footnotesize
$^{a}$ Truncated energy. For more exact value see Table~\ref{Cr_levels_table}\\
\indent $^{b}$ $\lambda_{\mathrm{air}}$ is calculated from the wavenumber using the modified \citet{Edlen66} dispersion formula by \citet{Birch94} for standard air\\
\indent $^{c}$ Transition probabilities taken from the calculations by \citet{Quinet97}
}
}
}

\begin{table}
 \caption{Ritz wavelengths of prominent parity-forbidden lines of \ion{Ti}{ii}, \ion{Cr}{ii} and \ion{Fe}{ii}, sorted by wavelength. From Tables~\ref{Fe_ritzterm_table}, \ref{Ti_ritzterm_table} and \ref{Cr_ritzterm_table}, which include the full set of lines with uncertainties and transition probabilities}
 \label{Strong_table}
 \centering
 \begin{tabular}{cr@{.}lr@{.}lrc}
 \hline\hline
 \\[-3mm]
 Ion & \multicolumn{2}{c}{$\lambda_{\mathrm{vac}}$} & \multicolumn{2}{c}{$\lambda_{\mathrm{air}}$} & Wavenumber & Multiplet\\
  & \multicolumn{2}{c}{(\AA )} & \multicolumn{2}{c}{(\AA )} & (cm$^{-1}$) & \\
 \hline
 \\[-3mm]
 \ion{Fe}{ii} &   7639&61670  &   7637&51419  & 13089.6619  & a$^{6}$D--a$^{4}$P \\
 \ion{Fe}{ii} &   7667&38923  &   7665&27922  & 13042.2491  & a$^{6}$D--a$^{4}$P \\
 \ion{Fe}{ii} &   7689&04345  &   7686&92760  & 13005.5189  & a$^{6}$D--a$^{4}$P \\
 \ion{Cr}{ii} &   8002&26949  &   8000&06908  & 12496.4549  & a$^{6}$S--a$^{6}$D \\
 \ion{Cr}{ii} &   8127&55768  &   8125&32343  & 12303.8192  & a$^{6}$S--a$^{6}$D \\
 \ion{Cr}{ii} &   8231&96395  &   8229&70150  & 12147.7694  & a$^{6}$S--a$^{6}$D \\
 \ion{Cr}{ii} &   8310&79498  &   8308&51122  & 12032.5433  & a$^{6}$S--a$^{6}$D \\
 \ion{Cr}{ii} &   8359&98415  &   8357&68710  & 11961.7452  & a$^{6}$S--a$^{6}$D \\
 \ion{Fe}{ii} &   8619&31631  &   8616&94915  & 11601.8483  & a$^{4}$F--a$^{4}$P \\
 \ion{Fe}{ii} &   8894&37171  &   8891&93015  & 11243.0651  & a$^{4}$F--a$^{4}$P \\
 \ion{Fe}{ii} &   9035&9708   &   9033&4909   & 11066.8795  & a$^{4}$F--a$^{4}$P \\
 \ion{Fe}{ii} &   9054&43465  &   9051&94977  & 11044.3119  & a$^{4}$F--a$^{4}$P \\
 \ion{Fe}{ii} &   9229&16055  &   9226&62838  & 10835.2216  & a$^{4}$F--a$^{4}$P \\
 \ion{Fe}{ii} &   9270&0992   &   9267&5560   & 10787.3710  & a$^{4}$F--a$^{4}$P \\
 \ion{Fe}{ii} &   9473&5428   &   9470&9444   & 10555.7131  & a$^{4}$F--a$^{4}$P \\
 \ion{Fe}{ii} &  12570&2064   &  12566&7681   &  7955.3189  & a$^{6}$D--a$^{4}$D \\
 \ion{Fe}{ii} &  12706&9109   &  12703&4355   &  7869.7333  & a$^{6}$D--a$^{4}$D \\
 \ion{Fe}{ii} &  12791&2242   &  12787&7260   &  7817.8600  & a$^{6}$D--a$^{4}$D \\
 \ion{Fe}{ii} &  15338&9391   &  15334&7484   &  6519.3557  & a$^{4}$F--a$^{4}$D \\
 \ion{Fe}{ii} &  15999&1453   &  15994&7751   &  6250.3339  & a$^{4}$F--a$^{4}$D \\
 \ion{Fe}{ii} &  16440&0180   &  16435&5279   &  6082.7184  & a$^{4}$F--a$^{4}$D \\
 \ion{Fe}{ii} &  16642&2527   &  16637&7076   &  6008.8019  & a$^{4}$F--a$^{4}$D \\
 \ion{Fe}{ii} &  16773&4034   &  16768&8226   &  5961.8193  & a$^{4}$F--a$^{4}$D \\
 \ion{Fe}{ii} &  17454&1593   &  17449&3933   &  5729.2934  & a$^{4}$F--a$^{4}$D \\
 \ion{Fe}{ii} &  17488&9900   &  17484&2146   &  5717.8831  & a$^{4}$D--a$^{4}$P \\
 \ion{Fe}{ii} &  17975&9706   &  17971&0627   &  5562.9820  & a$^{4}$F--a$^{4}$D \\
 \ion{Fe}{ii} &  18118&7982   &  18113&8514   &  5519.1299  & a$^{4}$D--a$^{4}$P \\
 \ion{Fe}{ii} &  18139&2612   &  18134&3089   &  5512.9037  & a$^{4}$D--a$^{4}$P \\
 \ion{Ti}{ii} & 100680&350    & 100652&908    &   993.2425  & a$^{4}$F--b$^{4}$F \\
 \ion{Ti}{ii} & 100996&941    & 100969&413    &   990.1290  & a$^{4}$F--b$^{4}$F \\
 \ion{Ti}{ii} & 101635&038    & 101607&335    &   983.9127  & a$^{4}$F--b$^{4}$F \\
 \ion{Ti}{ii} & 110136&46     & 110106&44     &   907.9646  & a$^{4}$F--b$^{4}$F \\
 \ion{Ti}{ii} & 112384&535    & 112353&903    &   889.8021  & a$^{4}$F--b$^{4}$F \\
 \ion{Ti}{ii} & 116056&188    & 116024&555    &   861.6516  & a$^{4}$F--b$^{4}$F \\
 \ion{Ti}{ii} & 121597&375    & 121564&232    &   822.3862  & a$^{4}$F--b$^{4}$F \\
 \ion{Ti}{ii} & 122872&15     & 122838&66     &   813.8540  & a$^{4}$F--b$^{4}$F \\
 \ion{Ti}{ii} & 131889&36     & 131853&41     &   758.2113  & a$^{4}$F--b$^{4}$F \\
 \ion{Ti}{ii} & 144111&16     & 144071&88     &   693.9088  & a$^{4}$F--b$^{4}$F \\
 \hline
 \end{tabular}
\end{table}

\section{Conclusions}

Data for forbidden lines in astrophysics are not easy accessible, in general, and for the infrared region, in particular. As forbidden lines by definition are associated with parity-forbidden transitions between low levels, many of the strongest lines appear in the near-infrared (transitions between LS-terms) and in the far-infrared (fine-structure transitions) in complex spectra. They are of great importance in nebular spectroscopy for studies of chemistry, temperature, density and dynamics of nebular clouds and regions.

To meet the requirements of accurate atomic data in the infrared region set by current and future high-resolution spectroscopy we have presented new wavelength measurements for forbidden lines of [\ion{Fe}{ii}], [\ion{Cr}{ii}] and [\ion{Fe}{ii}]. The data are based on revised energy level values obtained from very accurate wavelengths of allowed transitions in the three spectra studied.

\begin{acknowledgements}
We thank K.J.~\"Oberg for the use of his energy levels fitting routine and for discussions regarding this.
\end{acknowledgements}

\bibliographystyle{aa}

\end{document}